\documentclass[useAMS,usenatbib,usegraphicx]{mn2e}

\usepackage{times}

\newcommand{\rsun}{\,\mbox{$R_{\odot}$}}

\newcommand{\kms}{\hbox{km s$^{-1}$}}

\newcommand{\vsini}{\hbox{$v$\,sin\,$i$}}

\newcommand{\degs}{$\degr$}
\newcommand{\chisq}{$\chi^{2}$}

\newcommand{\ha}{H$\alpha$}

\newcommand{\aper}{$\alpha$ Persei}

\begin{document}

\title[Magnetic maps of HD 155555]{The first magnetic maps of a pre-main sequence binary star system - HD 155555}

 
\makeatother

\author[N.J.~Dunstone et al.]
{N.J.~Dunstone,$^1$\thanks{E-mail: njd2@st-andrews.ac.uk} G.A.J. Hussain,$^{1,2}$ A. Collier Cameron,$^1$ S.C. Marsden,$^3$ M. Jardine,$^1$
\newauthor
H.C. Stempels,$^1$ J.C. Ramirez Vélez,$^4$ J.-F.~Donati$^5$ \\
$^1$ School of Physics and Astronomy, University of St Andrews, Fife KY16 9SS, UK\\
$^2$ ESO, Karl-Schwarzschild-Strasse 2, D-85748 Garching, Germany \\
$^3$ Anglo-Australian Observatory, PO Box 296, Epping, NSW 1710, Australia \\
$^4$ LESIA, Observatoire de Meudon, 92195 Meudon, France \\
$^5$ LATT, CNRS-UMR 5572, Obs. Midi-Pyrénées, 14 Av. E. Belin, F-31400 Toulouse, France\\}

\date{2008, 200?}

\maketitle

\begin{abstract}

We present the first maps of the surface magnetic fields of a pre-main sequence binary system. Spectropolarimetric observations of the young, 18 Myr, HD 155555 (V824 Ara, G5IV + K0IV) system were obtained at the Anglo-Australian Telescope in 2004 and 2007. Both datasets are analysed using a new binary Zeeman Doppler imaging (ZDI) code. This allows us to simultaneously model the contribution of each component to the observed circularly polarised spectra. Stellar brightness maps are also produced for HD 155555 and compared to previous Doppler images.

Our radial magnetic maps reveal a complex surface magnetic topology with mixed polarities at all latitudes. We find rings of azimuthal field on both stars, most of which are found to be non-axisymmetric with the stellar rotational axis. We also examine the field strength and the relative fraction of magnetic energy stored in the radial and azimuthal field components at both epochs. A marked weakening of the field strength of the secondary star is observed between the 2004 and 2007 epochs. This is accompanied by an apparent shift in the location of magnetic energy from the azimuthal to radial field. We suggest that this could be indicative of a magnetic activity cycle. We use the radial magnetic maps to extrapolate the coronal field (by assuming a potential field) for each star individually - at present ignoring any possible interaction. The secondary star is found to exhibit an extreme tilt ($\approx$75\degr) of its large scale magnetic field to that of its rotation axis for both epochs. The field complexity that is apparent in the surface maps persists out to a significant fraction of the binary separation. Any interaction between the fields of the two stars is therefore likely to be complex also. Modelling this would require a full binary field extrapolation.

\end{abstract}

\begin{keywords}
Stars: pre-main sequence  --
Stars: magnetic fields --
(Stars:) binaries: spectroscopic --
Polarization --
Stars: imaging --
Stars: coronae --
\end{keywords}

\section{INTRODUCTION}

In the past decade our understanding of the surface magnetic topologies of stars beyond our Sun has greatly increased. This is mostly due to the application of the Zeeman-Doppler imaging (ZDI) technique (\citealt{semel89}) which allows us not only to map magnetic regions on stellar surfaces, but also to determine the orientation of the field lines in these regions. From combining observations in circularly polarised light with the positional information (using Doppler imaging principles), we can determine whether a particular region on the stellar surface is composed of mainly radial or azimuthal field. ZDI has now been applied to many rapidly rotating active stars with some surprising results (see \citealt{donati07} for a recent review).

The magnetic maps of active G and K dwarfs show regions of strong surface azimuthal field. For many of these targets the recovered azimuthal field strengths are in excess of the radial field (e.g. for AB Dor and LQ Hya, \citealt{donati03}), thus suggesting that the magnetic energy is concentrated in the toroidal field rather than the polodial field. This is quite different from what we find on the Sun where most of the resolved magnetic structures are radially orientated field lines. Current solar dynamo theory tells us that strong azimuthal field should only be found at the shear layer between the radiative core and convective envelopes (the tachocline). \cite{donati99ab} therefore suggest that the observation of strong surface azimuthal field on these rapidly rotating stars leads to the conclusion that the $\Omega$ dynamo component is not confined to the tachocline but instead may be distributed throughout the convective envelope.

The only tidally locked binary star to date for which published magnetic maps are available is the evolved primary star of the RS CVn system HR 1099. A strong latitudinal dependence on the polarity of the radial field maps is found and HR 1099 often exhibits a unipolar cap (\citealt{donati03}). Also present are strong, and often complete, axisymmetric rings of azimuthal field on the stellar surface. \cite{petit04} also confirmed the existence of an anti-correlation between the polarity of the radial and azimuthal fields. This is not observed on young rapidly rotating stars but is present on the Sun. Due to the evolved nature of RS CVn primary components, it is difficult to interpret their magnetic maps with respect to those of young rapidly rotating single stars. Therefore in an attempt to establish the relative importance of tidal locking and youth on surface magnetic field distributions, we present observations of the young pre-main sequence binary system HD 155555. 

The HD 155555 (V824 Ara) system is composed of a G5 IV primary and a K0 IV secondary and has an orbital period of 1.68 d. This makes it a good target for observing over a five day observing run, as outlined in \S \ref{sect:obs}. HD 155555 was first discovered as a close binary system by \cite{bennett67}. It was later classified as an RS CVn binary (\citealt{strass93}) based upon the activity and binarity of the system. \cite{pasquini91} suggested that HD 155555 was more likely a young pre-main sequence binary. This was based upon the high Li (6708 \AA) abundance (re-visited here in \S \ref{sect:atmo}) and the presence of a very active dMe companion (LDS587B). More recently \cite{strass00} derived an age of 18 Myr using the {\it{Hipparcos}} distance of 31 pc and the pre-main-sequence evolutionary tracks of \cite{d.antona97}. 

HD 155555 is a particularly active system with an X-ray luminosity of $2.7\rm{x10}^{30}\rm{erg\ s^{-1}}$ (\citealt{dempsey93}), both components show CaII H \& K core emission and filled in \ha. Therefore given its proximity and activity it is unsurprising that HD 155555 has been the subject of a number of recent papers. In 1996 it was the target of a multi-wavelength study by \cite{dempsey01}. The authors detected considerable flare activity, with several flares being observed simultaneously in the UV and radio wavelengths. As part of this campaign, optical spectroscopy was also obtained and surface brightness images were produced using Doppler imaging by \cite{strass00}. An earlier Doppler image of HD 155555 produced from data obtained in 1990 was published by \cite{hatzes99}. In addition, a magnetic field detection has also been reported for both components of HD 155555 by \cite{donati97survey}. 

In this paper we present new surface brightness maps (\S \ref{sect:spots}) and the first magnetic maps of HD 155555 (\S \ref{sect:magmaps}). To achieve this we have developed a binary ZDI code in order to model the contribution of each star to the combined Stokes V profiles that are observed during conjunction phases. We outline the details of this in section \S \ref{sect:zdi}. The more recent, and numerous, 2007 observations of HD 155555 are described first. Then in \S \ref{sect:2004} we present a smaller set of observations take in 2004. In obtaining magnetic maps for both components of a binary system we open up many interesting possibilities to explore the effects of binarity on the magnetic fields of both stars. Possible interaction between the stellar fields are of interest because it will determine many of the systems X-ray properties and the location of stellar winds. In \S \ref{sect:corextrap} we use the magnetic maps recovered of both the primary and secondary components in order to perform an initial analysis of the likelihood of interaction between the two stars. We discuss our results in \S \ref{sect:disc} and present our conclusions in \S \ref{sect:conc}.

\section{Observations and data reduction}
\protect\label{sect:obs}

Spectropolarimetric observations of HD 155555 were made at the Anglo-Australian Telescope (AAT) using the University College London \'{E}chelle Spectrograph (UCLES) which is fibre fed by the SemelPol visiting polarimeter (\citealt{semel93}) mounted at the Cassegrain focus. In total eleven uninterrupted nights of observations were secured on our target star HD 155555 in 2007 from March 30 to April 09. In this paper we only use data collected on five of these nights, from March 31 to April 04. The remaining six nights of observations are used in a forthcoming publication to study the surface rotation properties of both stars. The 1.68 d orbital period means that good phase coverage can be obtained in five nights, corresponding very closely to three stellar orbits.

The instrumental set-up is the same as that used on many previous Zeeman Doppler imaging runs and is described in detail by previous authors (\citealt{semel93}, \citealt{donati03}). Using SemelPol both left- and right-hand circular polarised light are captured simultaneously on the EEV2 CCD detector. Using the 31 gr mm$^{-1}$ grating full wavelength coverage between 4377 and 6819 \AA\ is achieved. A resolution 70,000 (i.e. 4.3 \kms) is obtained using this set-up. In order to supress all spurious polarisation signals (down to first order) due to instrumental drifts or temporal variations in stellar spectra (e.g. resulting from the orbital motions of the components in a binary system), circular polarisation (Stokes V) observations are made from a sequence of four exposures. Between each exposure the quarter-waveplate in SemelPol is rotated by +/- 45\degs, causing a switch of the circular polarisation carried in each fibre. The four sub-exposures are then combined using optimal extraction techniques as implemented by the Echelle Spectra Reduction: an Interactive Tool (ESpRIT) software developed by \cite{donati97survey}. In the process the normal calibrations pertaining to spectral data are performed and the spectra are wavelength calibrated. As HD 155555 is such a bright target (V = 6.9 mag) we were able to use 200 s exposures which meant that, after readout time, the total time taken for each sequence of four exposures (and so each Stokes I and V spectrum) was approximately 15 minutes.

In total 53 spectra were obtained over the five nights and are fairly evenly spread over the orbital phase of the binary. The signal-to-noise (S/N) ratio of both our Stokes I and V spectra in the peak order vary with the weather conditions between 70 and 230 with an average of around 160 per 1.9 \kms\ pixel. In order to greatly boost the signal-to-noise the technique of Least-Squares Deconvolution (LSD) is used as described by \cite{donati97survey} and implemented in ESpRIT. This effectively combines the absorption line profiles of many thousands of photospheric lines, resulting in a single high S/N ratio line profile. For HD 155555 we use a line-list for a K0 star produced by the VALD database (\citealt{vald95}, \citealt{vald99}). All strong chromospheric lines (such as the Balmer series) are excluded. Approximately 2800 lines are used for LSD which result in a S/N ranging from 2500 to 10,500 for Stokes V spectra and between 680 and 1200 for Stokes I spectra. We take advantage of the strong telluric lines present in the spectra and the LSD technique in order to correct for drifts in our wavelength calibration over the course of the night. \cite{donati03} show that this calibration results in radial velocities which are stable to better than 0.1 \kms.

\section{System parameters and brightness maps}
\protect\label{sect:spots}

We use the Doppler imaging code `DoTS' (\citealt{cam97dots}) to map the surface brightness distribution of the surfaces of both stars using the Stokes I intensity spectra. This code uses maximum entropy, as implemented by \cite{skilling84}, to return the simplest possible surface map at a given level of \chisq. A two-temperature model is used to describe the surface of each star (\citealt{cam94twotemp}). Spectroscopic look up tables are created to model the contribution from each element on the stellar surface. Slowly rotating reference stars are fitted with Gaussian profiles which we use to model both the local intensity profiles of the photosphere and the spot, for each star. It has been shown by \cite{unruh95} that using synthetic Gaussian profiles instead of those from slowly rotating stars has little effect on the resulting brightness maps. While the obvious advantage of using synthetic Gaussians is that they are noiseless. For the primary star we use temperatures of 5300 K and 4050 K for the photosphere and spot respectively. Similarly the cooler secondary star is modelled using a photospheric temperature of 5050 K and a spot temperature of 3787 K. These temperatures come from the spectroscopic analysis described in \S \ref{sect:atmo}.

\subsection{Determining system orbital parameters}
\protect\label{sect:syspar}

\begin{table}
\caption{Orbital and physical parameters for the HD 155555 system.}
\protect\label{tab:syspar}
\begin{center}
\begin{tabular}{cccc}
\hline
Element & Unit & Value & Source \\
\hline	
Orbital: \\
$P$		& (days) 	& 1.6816463			& adopted$^{1}$ \\
$T_{0}$		& (HJD) 	& 2446997.9102			& adopted$^{1}$ \\
$\Phi_{0}$	& 		& 0.752474			& DoTS \\
$e$		&  		& 0.0 		 		& adopted \\
$\gamma$	& (\kms) 	& 3.72$\pm$0.02  		& DoTS \\
$q$	& $m_{2}/m_{1}$	& 0.935$\pm$0.001  		& DoTS \\
$K_{1}$		& (\kms) 	& 86.4$\pm$0.1	 		& DoTS \\
$K_{2}$		& (\kms)	& 94.7$\pm$0.2  		& ($K_{1}$ and $q$) \\
$i$		& (degs)	& 50 (52 adopted)	& DoTS \\
$(v \sin i)_1$  & (\kms)        & 34.9				& DoTS \\
$(v \sin i)_2$  & (\kms)        & 31.3				& DoTS \\
\multicolumn{2}{c}{Physical (assuming $i=52$\degs):}  & &\\
$m_{1}$		& ($M_{\odot}$)	& 1.054				& DoTS \\
$m_{2}$		& ($M_{\odot}$)	& 0.986				& DoTS \\
$R_{1}$		& ($R_{\odot}$)	& 1.47				& DoTS \\
$R_{2}$		& ($R_{\odot}$)	& 1.32				& DoTS \\
$T_{{\rm{eff}}, 1}$	& (K)		& 5300$\pm$100			& SME \\
$T_{{\rm{eff}}, 2}$	& (K)		& 5050$\pm$100			& SME \\
${\rm{log}}\ g_{1}$&	 	& 4.05$\pm$0.1			& SME \\
${\rm{log}}\ g_{2}$& 		& 4.10$\pm$0.1			& SME \\
\hline
\multicolumn{4}{l}{$^{1}$~from \citet{strass00}.}  \\
\end{tabular}
\end{center}
\end{table}

In addition to producing surface brightness maps we can use DoTS to determine the fundamental parameters of the HD 155555 system. This is achieved by finding the values of each system parameter that give the best fit to the stellar profiles (i.e. that minimise \chisq) after a fixed number of maximum entropy iterations {{(see \citealt{barnes00})}}. In a binary system the additional orbital parameters greatly increase the computer time required to achieve a global solution. We approached this problem by carefully selecting combinations of two parameters to be minimised for simultaneously. In practice we used a pool of 30 computers to simultaneously run different combinations of the variables. Several iterations were made through the pairs of parameters until a global convergence on a solution with \chisq\ = 0.6 was achieved. {{At this point further changes in the parameters were of the order of their uncertainties as estimated from the curvature of the \chisq\ surfaces}}. Values for system parameters are collated in Table \ref{tab:syspar}. 

We use the ephemeris of \cite{strass00} which is based on over twenty years of radial velocity data. We attempted to derive our own value for the period of the system using our 2007 dataset alone and found that it was consistent with that of \cite{strass00}, however with naturally larger uncertainty. The value of 0.752474 we obtain for the phase offset ($\Phi_{0}$) reflects the traditional eclipsing binary definition that we use in DoTS. Here phase zero occurs at stellar conjunction with the primary star furthest away from the observer (i.e. normally primary eclipse in an eclipsing binary system). We make the simplifying assumption that both components of HD 155555 have rotational axes perpendicular to the orbital plane. Unfortunately determining the inclination of the binary orbit for non-eclipsing systems is as problematic as the rotation axis of single stars because the Doppler imaging process is only weakly sensitive to this parameter. We derive a value of 50\degs\ (using the \chisq\ minimisation technique described above) which is consistent with the value of 52\degs\ obtained by both \cite{strass00} and \cite{pasquini91}. For consistency with these works we therefore also adopt a value of 52\degs.

\subsection{Atmospheric and evolutionary parameters}
\protect\label{sect:atmo}

A subset of our highest signal-to-noise spectra are selected to determine the
atmospheric parameters using spectral synthesis. We apply the technique of
tomographic separation (see  \citealt{bagnuolo91}) to disentangle the two
components using the orbital parameters derived in the last section. We then
calculated and fitted synthetic spectra to each component with the help of the
Spectroscopy Made Easy (SME) code (\citealt{valenti96}). This code uses
least-squares minimization to determine the atmospheric parameters that best
describe the observed spectra. For these calculations we used stellar model
atmospheres from the \citet{kurucz93} grid, and atomic line data obtained from
the VALD database \citep{vald95,vald99}. In our calculations we assumed
solar abundances, a microtubulence of $1.75$~km\,s${}^{-1}$ and a 
macrotubulence
of $2.0$~km\,s${}^{-1}$, all typical values for a G5--K0 class 
pre-main-sequence
object.

For our analysis we selected two wavelength regions, one around the temperature
and gravity sensitive Na {\sc i} D lines, and one between $6000$ and $6200$
{\AA}, a region that contains a large number of metal lines. The line ratios in
the latter regions provide independent leverage on the effective temperatures.
We find the effective temperatures and gravities of the two stars to be very
similar (primary: $T_{\rm eff, 1} = 5300 \pm 100$ K and $\log g_1 = 4.05 \pm
0.1$, secondary $T_{\rm eff, 2} = 5050 \pm 100$ K and $\log g_2 = 4.10 \pm
0.1$). The full set of parameters is listed in Table \ref{tab:syspar}.

Within their uncertainties, the values we find are consistent with the values
quoted in \cite{pasquini91}. The main difference we find is that the surface
gravity of the secondary star is significantly lower than the value of $4.5$
quoted by \cite{strass00} and consequently much closer to the value obtained 
for
the primary star. This is also expected from the orbital parameters of the
system, since $g \sim {\rm G}M/R^{2}$ and thus: 
$$\Delta(\log g) = \log g_1 - \log g_2 = \log[M_1 / M_2 * (R_2 / R_1)^2]$$ Substituting $M_1 / M_2 = K_2 / K_1$ and assuming synchroneous rotation, 
$R_2 / R_1 = (v \sin i)_2 / (v \sin i)_1$ this gives:
$$\Delta(\log g) = \log (K_2 / K_1 * \left[(v \sin i)_2 / (v \sin i)_1\right]^{2})$$ 
Using the values from Table \ref{tab:syspar}, one finds $\Delta(\log g) = -0.05$.

We also re-examine the lithium abundance to provide an indication of youth.
Using the atmospheric parameters above, we find $\log n({\rm Li}) = 3.16$
(primary) and $3.15$ (secondary). Our analysis takes account of the non-LTE
corrections published by \cite{carlsson94} that were not available to
\cite{pasquini91}. These lithium abundances are slightly below the primordial
values of 3.2-3.3, and suggest a lithium age of 0--30 Myr 
(\citealt{sestito05}).
This is in good agreement with the age of 18 Myr derived by \cite{strass00} 
from
the pre-main sequence tracks of \cite{d.antona97}. We therefore agree with the
conclusions of \cite{pasquini91} and \cite{strass00} that HD 155555 is a young
pre-main sequence binary system and not an evolved RS  CVn object.

\subsection{Brightness images}

\begin{figure*}
 \begin{center}
  \includegraphics[width=12cm,angle=270]{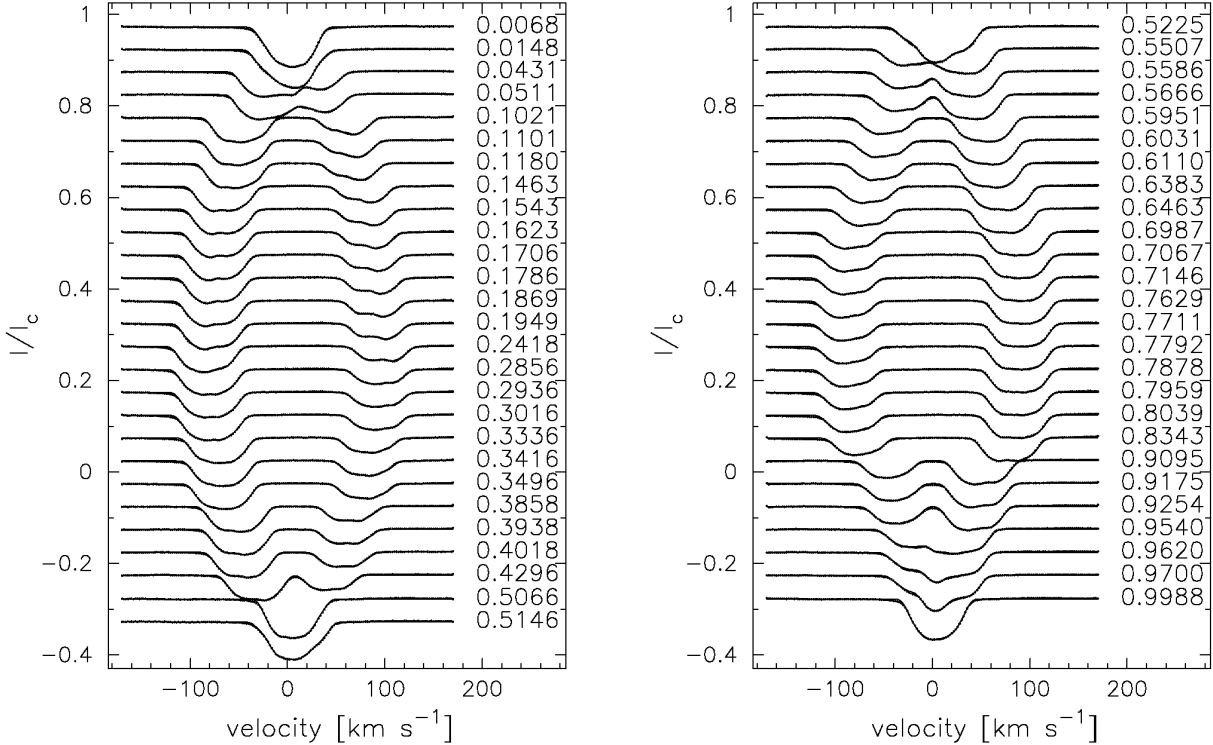}
 \end{center}
\caption[Stokes I fits]{The resulting 2007 LSD profiles are plotted along with their uncertainties. The solid line is the maximum entropy regularised fit we obtain to the data from Doppler imaging. Careful examination of the line profiles reveals the presence of spots, seen as raised bumps in the line profiles. Profiles are ordered by orbital phase (listed to the right of each profile) and clearly show the motion of both binary components. For reference, the primary star is the profile seen generally at negative velocities for phases $\phi=0.0-0.5$. Included in the on-line version of this paper is a plot showing the starspots in more detail.}
\protect\label{fig:fitspot}
\end{figure*}

Now that the best parameter set for HD 155555 has been found, we can attempt to model the deviations (bumps) in the line profile due to the presence of stellar spots. Unlike earlier in this section where we were interested in the parameter set that minimises \chisq, when generating final spot maps a target \chisq\ value is set such that the Maximum Entropy iterations converge before we start over-fitting the data and producing artificially noisy maps. In Fig. \ref{fig:fitspot} we show the 53 observed Stokes I LSD profiles and the corresponding fits we obtain using DoTS that correspond to a reduced \chisq = 0.7. A value for \chisq\ below unity is often achieved when performing Doppler imaging using intensity spectra (Stokes I), it simply indicates that the S/N values from the LSD process are slightly underestimated. The profiles have been ordered by phase so the binary motion can clearly be seen and individual spots are easily traced. From the LSD profiles alone we can see that the flat bottomed shape of the rotation profiles of both stars indicates the presence of polar spots.

\begin{figure*}
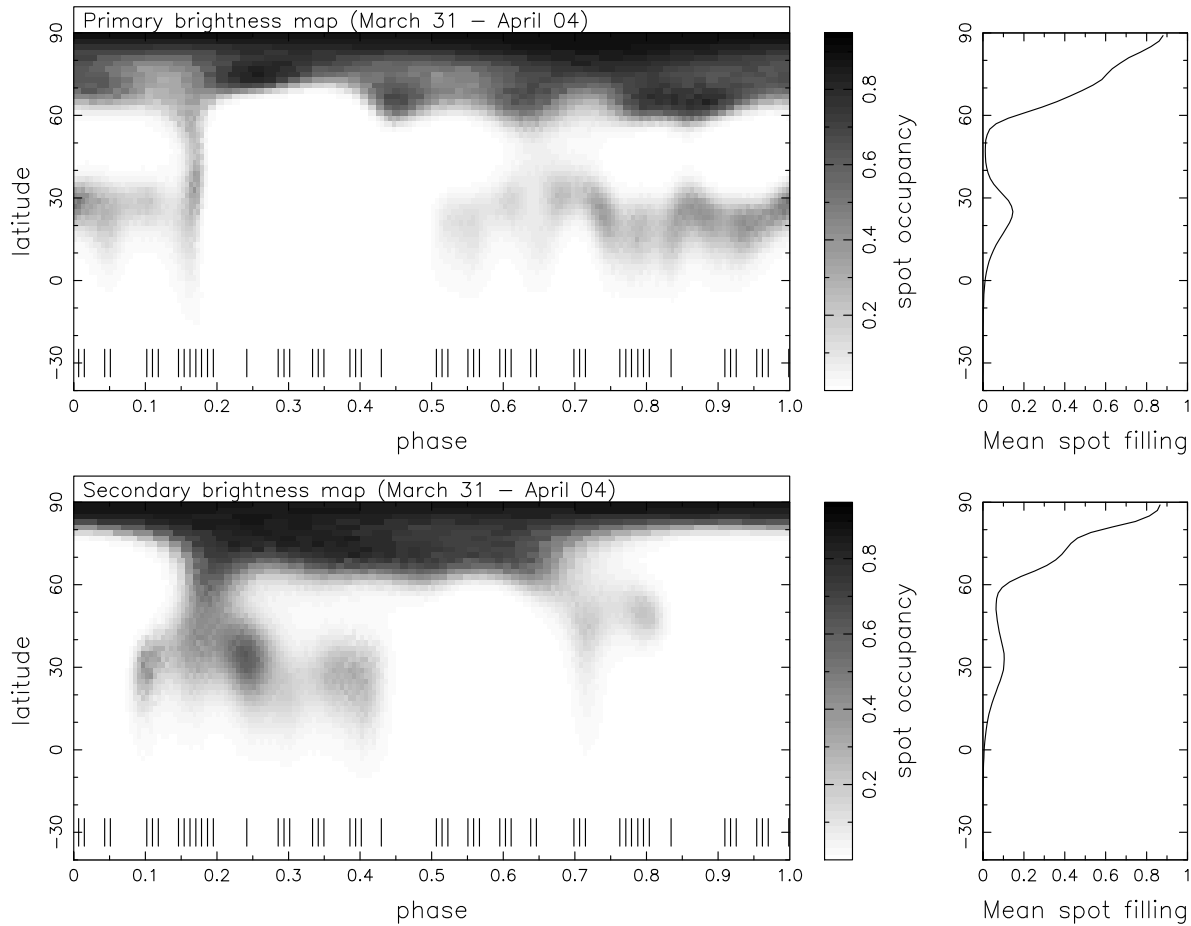

\begin{center}
  \begin{tabular}{cc}
    \includegraphics[width=6.0cm,angle=270]{Image/njd_fig2.ps} \\
    \includegraphics[width=6.0cm,angle=270]{Image/njd_fig3.ps} \\
  \end{tabular}
\end{center}
\caption[Brightness maps]{Doppler images of the surface brightness distribution of both components in 2007, showing the location of the prominent star spots. The tick marks at the bottom of each main panel mark the phases of observation. To the right of each map is a graph showing the mean spot filling factor as a function of stellar latitude. Note that for the primary star this is significantly bimodal.}
\protect\label{fig:spotmaps}
\end{figure*}

In Fig. \ref{fig:spotmaps} the brightness maps produced by the Doppler imaging process are shown. We choose to leave these maps labelled with phase, as opposed to the more conventional stellar longitude, so that the locations of spots can be more easily compared to the binary orbital phase. Note that for the primary star phase $\phi = 0.0$ faces the secondary star, while for the secondary star phase $\phi = 0.5$ faces the primary. The first obvious characteristic of both images are the large polar spots. On the primary star individual darker spots can be made out at latitudes between 60\degs\ and the pole. A band of low-latitude spots between (10 - 40\degs) is also present and covers three quarters of the stellar circumference. While we have been able to resolve some structure in this band it is likely that in reality it is composed of smaller, unresolved spots.  Latitudes between 40 - 60\degs\ are remarkably devoid of spot coverage producing the bi-modal plot of mean spot filling factor versus stellar latitude shown in the right panel of Fig. \ref{fig:spotmaps}. 

\begin{figure*}
\begin{center}

  \begin{tabular}{cc}
    \includegraphics[width=6.cm,angle=270]{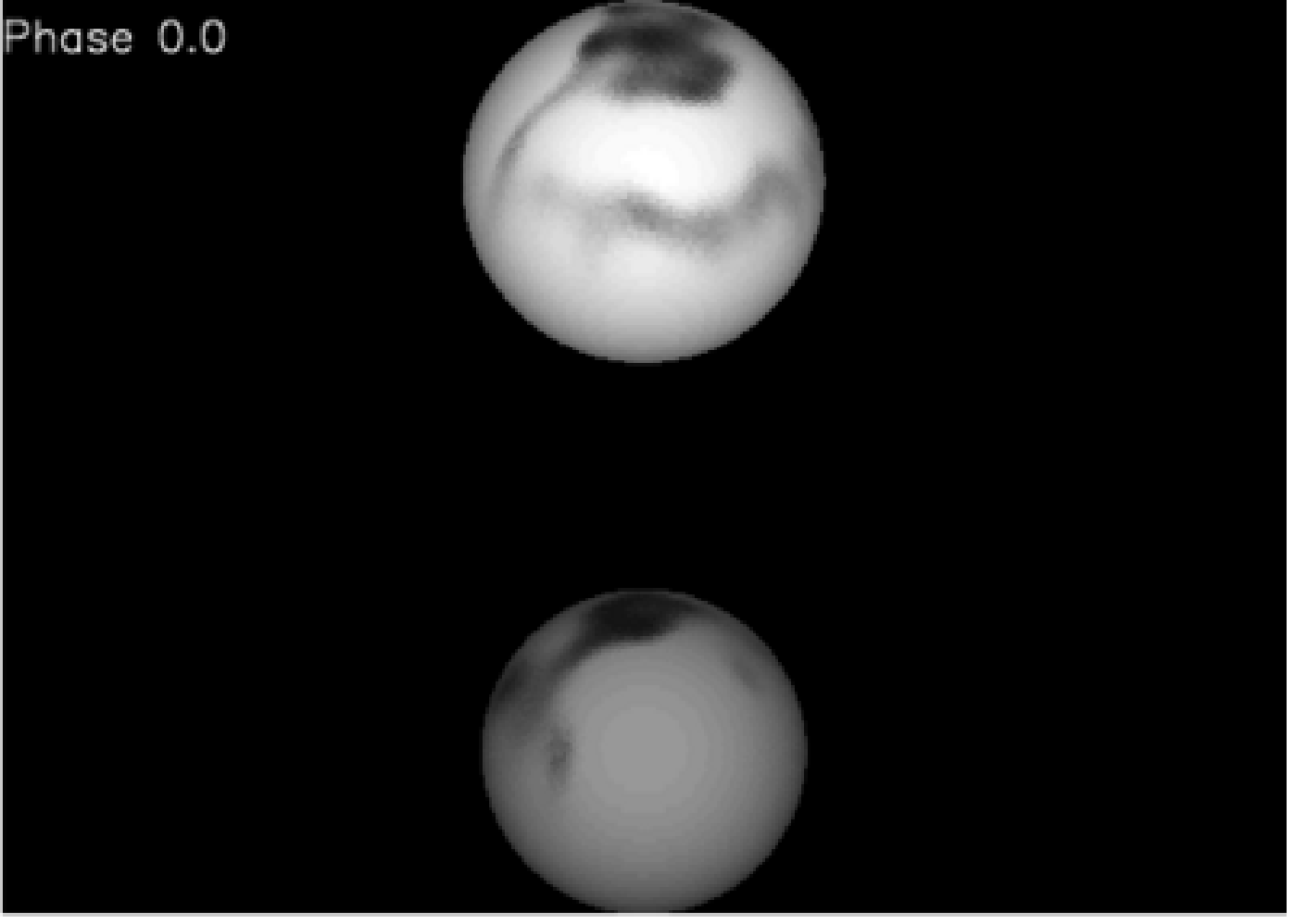} &
    \includegraphics[width=6.cm,angle=270]{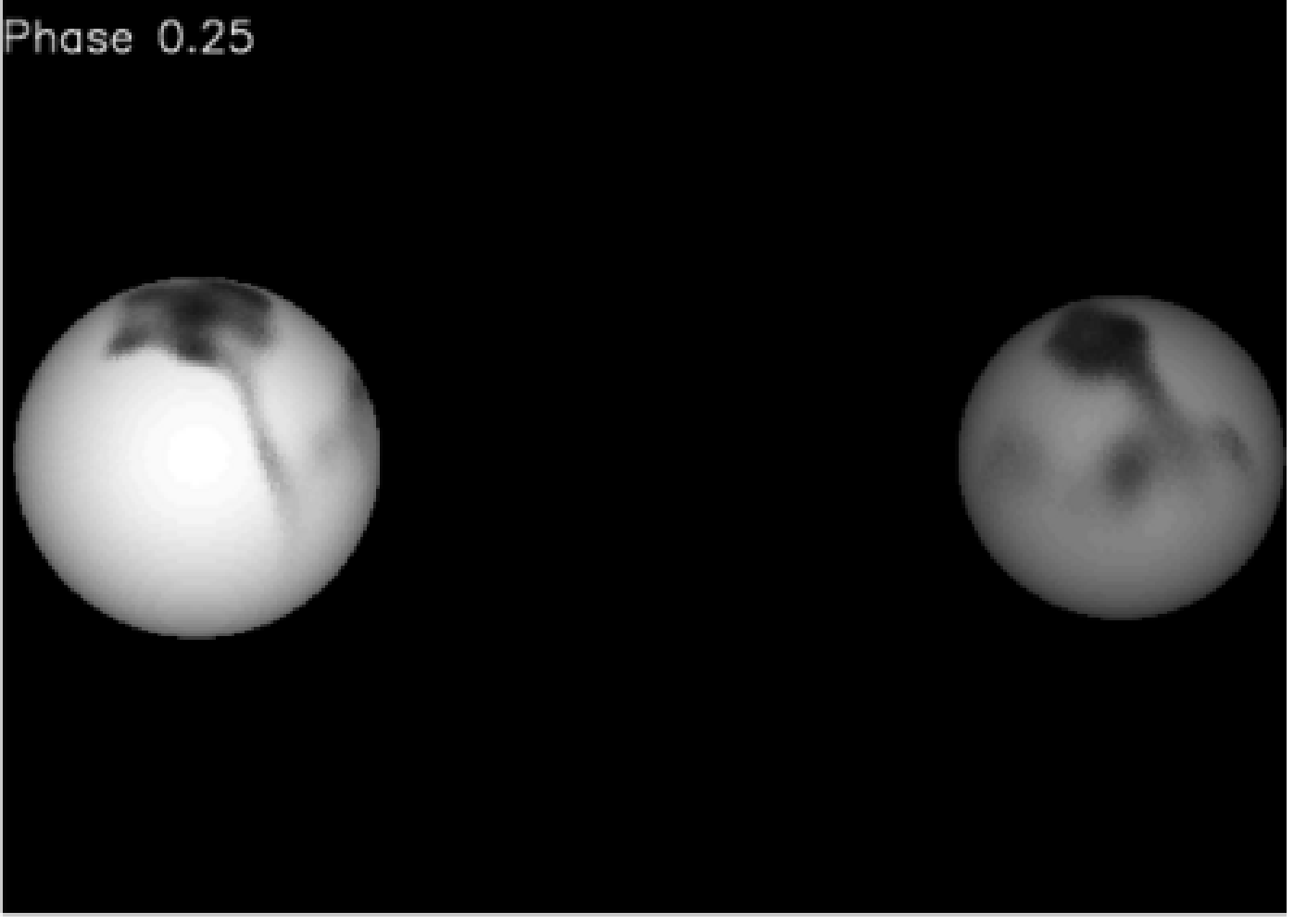} \\
    \includegraphics[width=6.cm,angle=270]{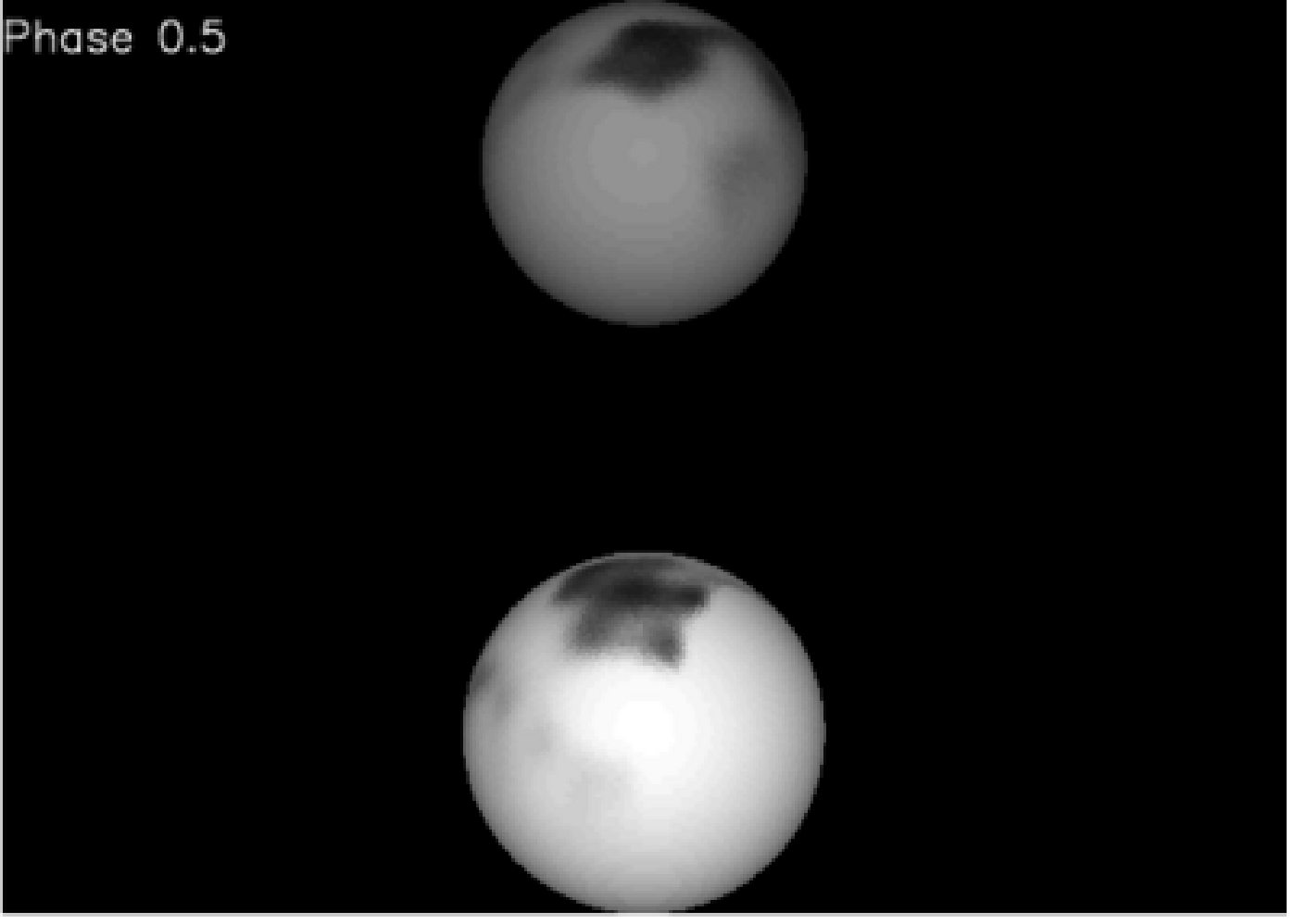} &
    \includegraphics[width=6.cm,angle=270]{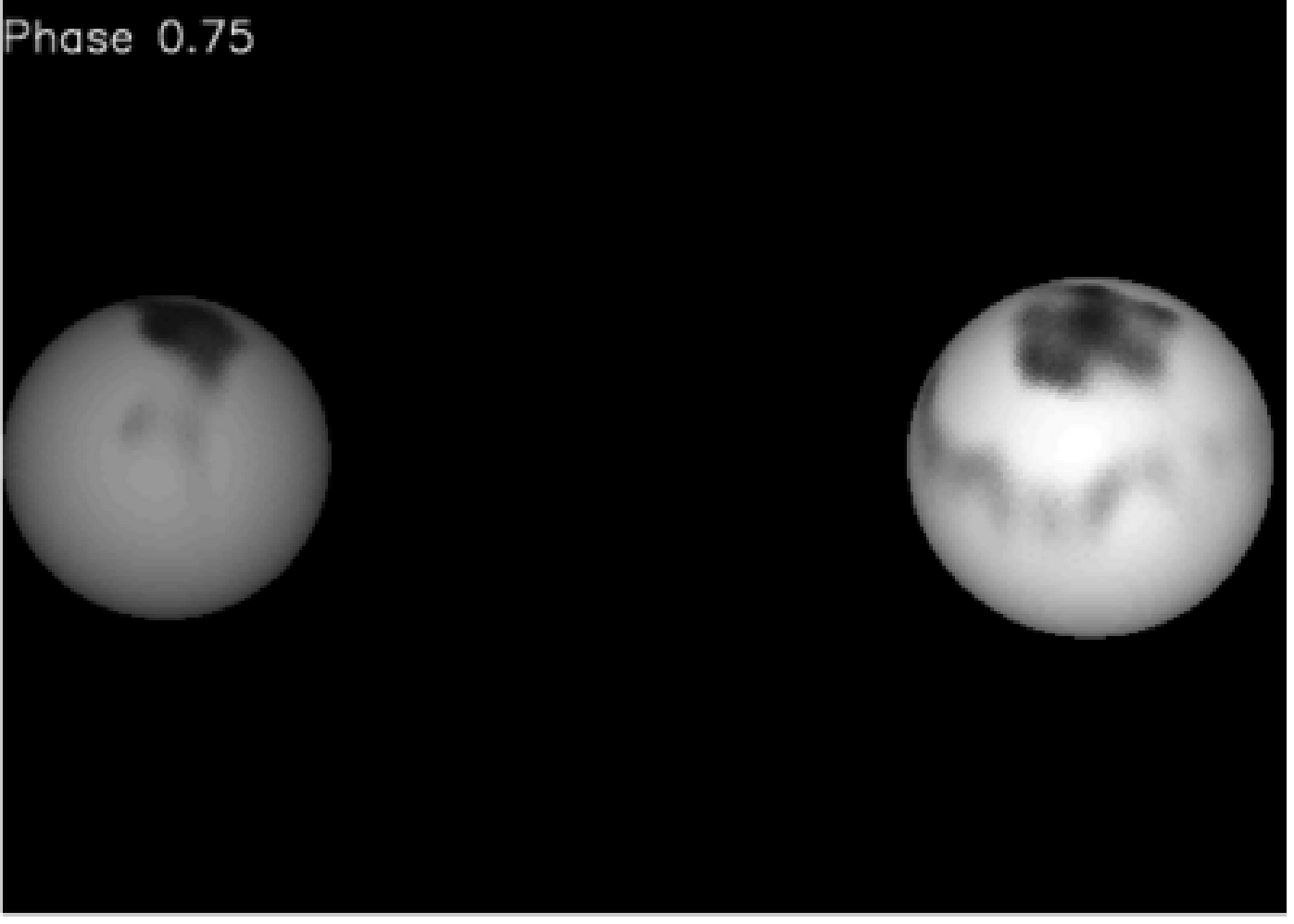} \\
  \end{tabular}

\end{center}
\caption[Brightness maps painted onto stellar surfaces]{The 2007 Doppler brightness images, in Fig. \ref{fig:spotmaps}, are shown projected onto the surfaces of each star at the four quadrature phases. The unspotted photosphere of the secondary star appears darker to the primary star due to the difference in stellar temperatures.}
\protect\label{fig:spotmovie}
\end{figure*}

The secondary star has a large offset polar spot that is centred at phase $\phi = 0.4$ and reaches down to a latitude of 60\degs. A large spot group at $\phi = 0.2$ is the lowest latitude feature reaching down to 20\degs\ and is clearly seen in the LSD profiles of Fig. \ref{fig:fitspot}. A weak pair of mid-latitude spots is also recovered at phases $\phi = 0.7$ and $\phi = 0.8$. In Fig. \ref{fig:spotmovie} we show the brightness maps projected on to the stellar surfaces so that the locations of spots can be compared to the binary orbital phase. We discuss the starspot distributions further in \S \ref{sect:disc} where we also compare them to previously published maps.

\section{Zeeman-Doppler imaging of binary systems}
\protect\label{sect:zdi}

All previous magnetic maps have been produced using ZDI codes designed to work on single stars. Therefore the few maps of the primary stars of binary systems that have been produced to date were obtained by performing the initial step of correcting the LSD profiles for the orbital motion. The data can then be treated as that of a single star. This is a common procedure for modelling of both Stokes I and V data by codes that are not capable of modelling the contribution from both stars. Due to the continuum present in Stokes I spectra this is particularly difficult. The binary conjunction phases are either not observed, or the part of the primary rotational profile that overlaps with the secondary star is masked out, or the secondary profile is removed by subtraction of a Gaussian.  The situation when using Stokes V data is considerably easier however, this is because the continuum is unpolarised. For RS CVn binaries the magnetic field strength of the secondary star is so small (due in part to the ratio of the stellar luminosities) that its contribution to the combined Stokes V signatures can often be assumed negligible. It is this assumption that is used to produce the many magnetic images of the primary star in the RS CVn binary system HR1099 (e.g. \citealt{petit04hr1099}).

In the case of the near equal mass ratio binary HD 155555, the stars are so similar that the above options are not viable. For approximately one third of the orbital period the rotation profiles of the stars overlap to some extent, as illustrated in Fig. \ref{fig:fitspot}. If one chose to simply not observe the conjunction phases then there would be two phase gaps each of duration $\Delta\phi \simeq 0.16$. These would at the very least reduce the level of detail recovered in the images and may even produce imaging artifacts. Gaps in the phase coverage is worse for magnetic than brightness imaging due to the fact that we ascertain the nature of the field (e.g. radial or azimuthal) by the behaviour of the Stokes V signature as it moves through the line profile. As we shall see, the magnetic field strengths on the two stars are comparable. Therefore the only efficient way to proceed is to develop a code capable of performing full ZDI of a binary star system.

We base our new ZDI code on that of \cite{hussain00} which itself was originally spawned from the DoTS code, we therefore name the new code `ZDoTS'. One of the main reasons why the binary version was not previously implemented was due to the computational demands that this would place. The memory requirements of ZDoTS exceeds that of the DoTS by a factor of three due to the necessity of storing each of the magnetic vectors (the radial, azimuthal and meridional fields). ZDoTS uses array compression and takes advantage of the general increase in recent computer memory in order to overcome this problem.

When the two binary components are in conjunction we model the result as a linear superposition of the two Stokes V signatures from both stars. Just as in Doppler imaging observations, such phases alone do not provide information about the relative contribution to the observed Stokes V spectrum from each star. However, when combined with observations where the two components are separated in velocity space then knowing how the two Stokes V signatures combine helps greatly to constrain the geometry of the magnetic field maps. 

\begin{figure}
 \begin{center}
  \includegraphics[width=8cm,angle=0]{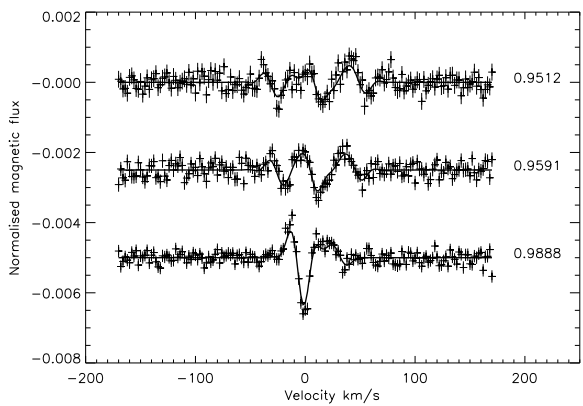}
 \end{center}
\caption[Illustrating Zeeman Doppler imaging in a binary system]{Three example Stokes V LSD profiles are shown to illustrate the ability of our code to reproduce the complex superposition of Stokes V signatures that arise when the stars are in conjunction. Note the two Stokes V minima (one from each star) that move closer together due to orbital motion between the top and middle spectra. They then combine in the third spectrum to produce the largest amplitude signal we observe during our campaign. We note that the signal-to-noise of the top two spectra shown here are below average.}
\protect\label{fig:together}
\end{figure}

As in the single star ZDI code of \cite{hussain00}, we model the local Stokes V line profiles using a Gaussian profile (which is assumed constant over the stellar surface) and we assume that the weak field approximation holds. Three consecutive Stokes V spectra of HD 155555, shown in Fig. \ref{fig:together}, illustrate particularly well what can happen when the stars are in conjunction. The Stokes V signatures from the two stars can be seen moving closer together between the top and bottom spectra of Fig. \ref{fig:together}. Then during the conjunction phase in the bottom spectrum we observe, by chance, the exact superposition of the troughs of the two profiles which produces a large amplitude combined signature. As Fig. \ref{fig:together} shows, the ZDoTS code has successfully modelled this. We note that the three spectra shown in Fig. \ref{fig:together} were not obtained in the five nights considered in this publication but from the subsequent five nights of data (specifically, April 05). The ZDoTS code was tested by creating synthetic Stokes V spectra from an input map with spots located at different

In order to test ZDoTS further we used the surface magnetic maps of the single, K0 dwarf, AB Dor (\vsini\ = 91 \kms) from 1996 and 1999 (\citealt{donati97ab} and \citealt{donati03}). We `painted' one of these maps on to the primary star and the other onto the secondary and assumed all the same orbital and physical parameters of the HD 155555 system (from Table \ref{tab:syspar}). {{Synthetic spectra were then generated, using the forward module of ZDoTS, of comparable signal-to-noise and at each of the same 53 observational phases as our real data. This resulted in Stokes V spectra of the binary system including the tangled conjunction phases. The spectra were then used as input into ZDoTS and we attempted to reconstruct the magnetic maps. The results were compared with the performance of the original (single star) ZDI code of \cite{hussain00} at recovering the magnetic maps of each star in turn. The results showed that both codes recovered essentially the same maps. When the images were collapsed in latitude the maximum difference between the two magnetic fluxes was always less than 5\%. Importantly, there was no visible phase (longitude) dependence on the accuracy of the reconstructed maps. This illustrates that ZDoTS is as capable of reproducing surface maps for binary systems as for single stars, at least for well phase sampled data. To examine further tests of ZDoTS using synthetic data see \cite{dunstone08thesis}.}}

\subsection{Magnetic maps of HD 155555}
\protect\label{sect:magmaps}

\begin{figure*}
 \begin{center}
  \includegraphics[width=10.5cm,angle=270]{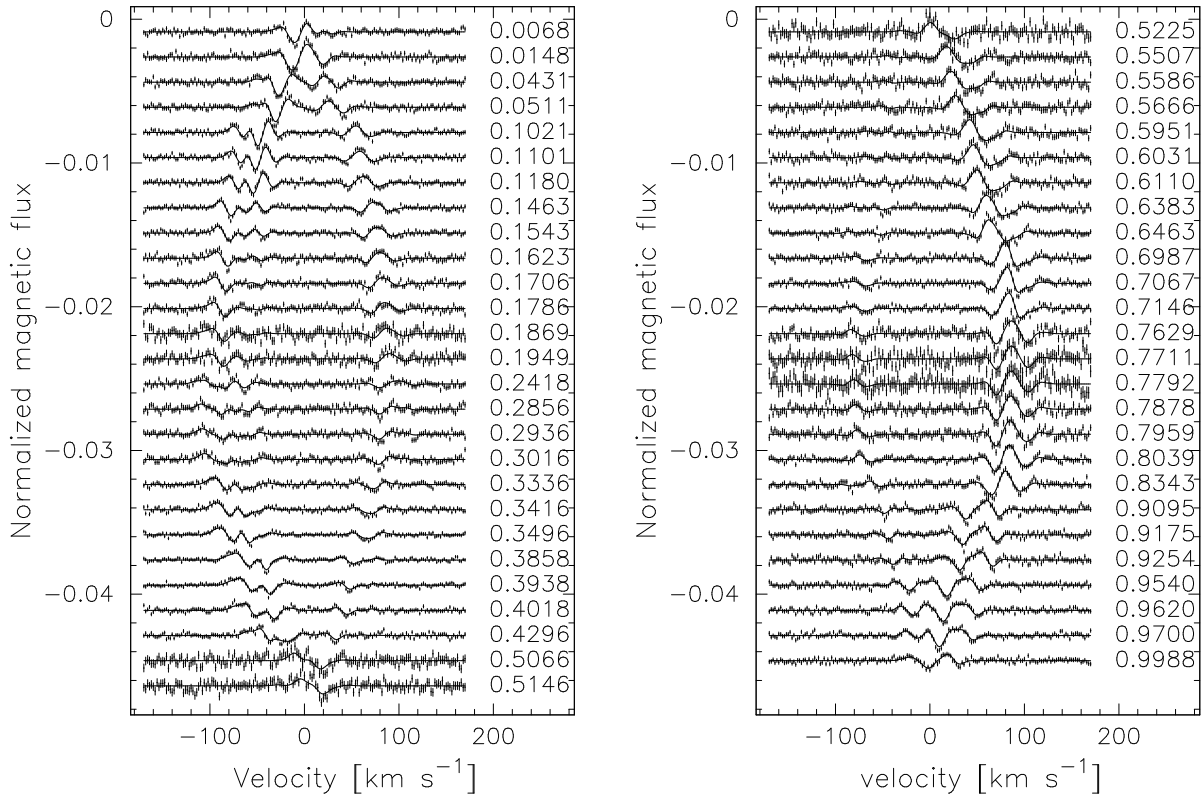}
 \end{center}
\caption[Stoke V fits]{The 2007 resulting observed Stokes V LSD profiles are plotted along with their uncertainties. The solid line is the maximum entropy regularised fit we obtain to the data from Zeeman Doppler imaging. Profiles are ordered by orbital phase (listed to the right of each profile) and clearly show the motion of both binary components.}
\protect\label{fig:fitmag}
\end{figure*}

In Fig. \ref{fig:fitmag} we show the 53 Stokes V LSD profiles along with the maximum entropy regularised fits corresponding to \chisq = 1.0. As in Fig. \ref{fig:fitspot} we order the Stokes V profiles by phase and again the binary orbital motion can be clearly seen. Unlike in Fig. \ref{fig:fitspot}, it is easy in Fig. \ref{fig:fitmag} to spot the few spectra which have substantially poorer signal-to-noise. These occur around phases $\phi=0.52$ and $\phi=0.77$ and still provide some constraint to modelling the surface magnetic field.

The strengths and complexity of both the primary and secondary Stokes V signatures varies with orbital phase. From a visual inspection of Fig. \ref{fig:fitmag} it appears from the Stokes V profiles that the primary star often possesses a more complex field with either more changes of polarity and/or different field orientations than the secondary star. This is illustrated by the greater number of sign reversals seen throughout the line profile of the primary star (as many as six seen around phase $\phi=0.15$). The Stokes V signature from the secondary star all but disappears around phase $\phi=0.6$.

\begin{figure*}
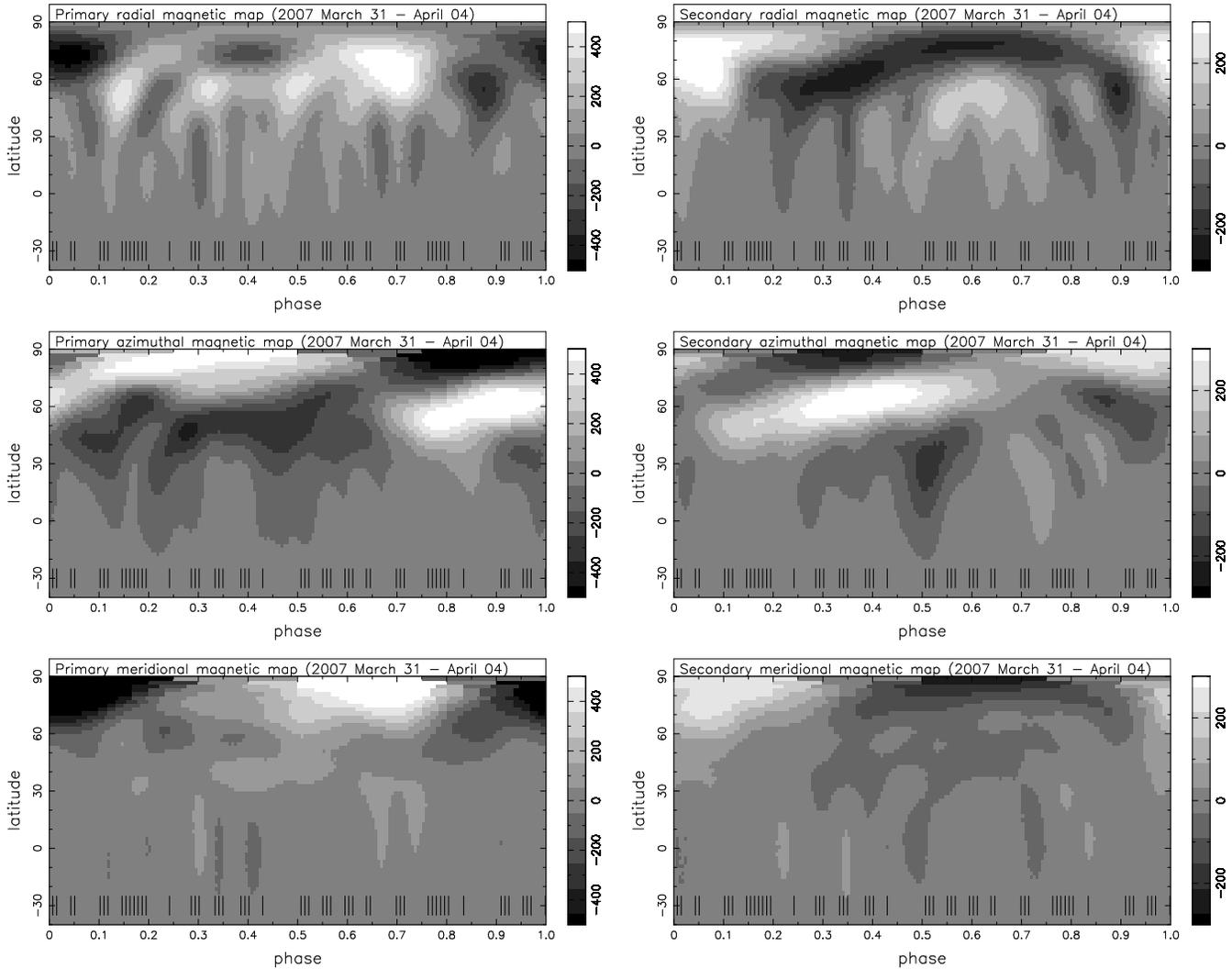

 \begin{center}
  \begin{tabular}{cc}
    \includegraphics[width=4.5cm,angle=270]{Image/njd_fig10.ps} &
    \includegraphics[width=4.5cm,angle=270]{Image/njd_fig11.ps} \\
    \includegraphics[width=4.5cm,angle=270]{Image/njd_fig12.ps} &
    \includegraphics[width=4.5cm,angle=270]{Image/njd_fig13.ps} \\
    \includegraphics[width=4.5cm,angle=270]{Image/njd_fig14.ps} &
    \includegraphics[width=4.5cm,angle=270]{Image/njd_fig15.ps} \\
  \end{tabular}
\caption[Magnetic maps]{The 2007 magnetic surface maps produced from the Zeeman Doppler imaging process. Maps of the primary star are shown on the left, while those of the secondary are on the right. The top panels show the radial magnetic field, the middle shows the azimuthal maps and the bottom panels show the meridional field. In all maps the neutral grey colour represents zero magnetic field, white shows positive polarity and black shows regions of negative polarity. Note that the greyscale contrast range is different between the primary and secondary maps, as illustrated in the key to the immediate right of each map.}
\protect\label{fig:magmaps}
\end{center}
\end{figure*}

The reconstructed magnetic images are shown in Fig. \ref{fig:magmaps}. We show the radial, azimuthal and meridional field components. The density and coverage of the observations ensures that there are no imaging artifacts that can result from poor phase sampling. The maximum amplitude of both the radial and azimuthal field is approximately twice as large on the primary as on the secondary star. Note that ZDI is only sensitive to high latitude meridional flux on stars with moderate to high inclination angles. At low latitudes the meridional map suffers considerable cross-talk with the radial map (\citealt{donati97recon}). A comparison of the recovered low latitude meridional field (bottom panels Fig. \ref{fig:magmaps}) with the radial field (top panels Fig. \ref{fig:magmaps}) shows this to be the case. 

The radial field map of the primary star shows many small regions of flux of alternating polarities. The level of surface detail recovered is in fact near the limit that our observations of a star with a relatively small \vsini\ ($\simeq35$\kms) could produce. As an interesting comparison, we note that the radial field maps of the primary star are similar in the level of detail they recover to those produced of AB Dor (actual \vsini=91\kms) once spun-down to \vsini=35\kms, as was performed during the testing of ZDoTS described earlier.

\begin{figure*}
 \begin{center}
  \begin{tabular}{cc}
    \includegraphics[width=6.cm,angle=270]{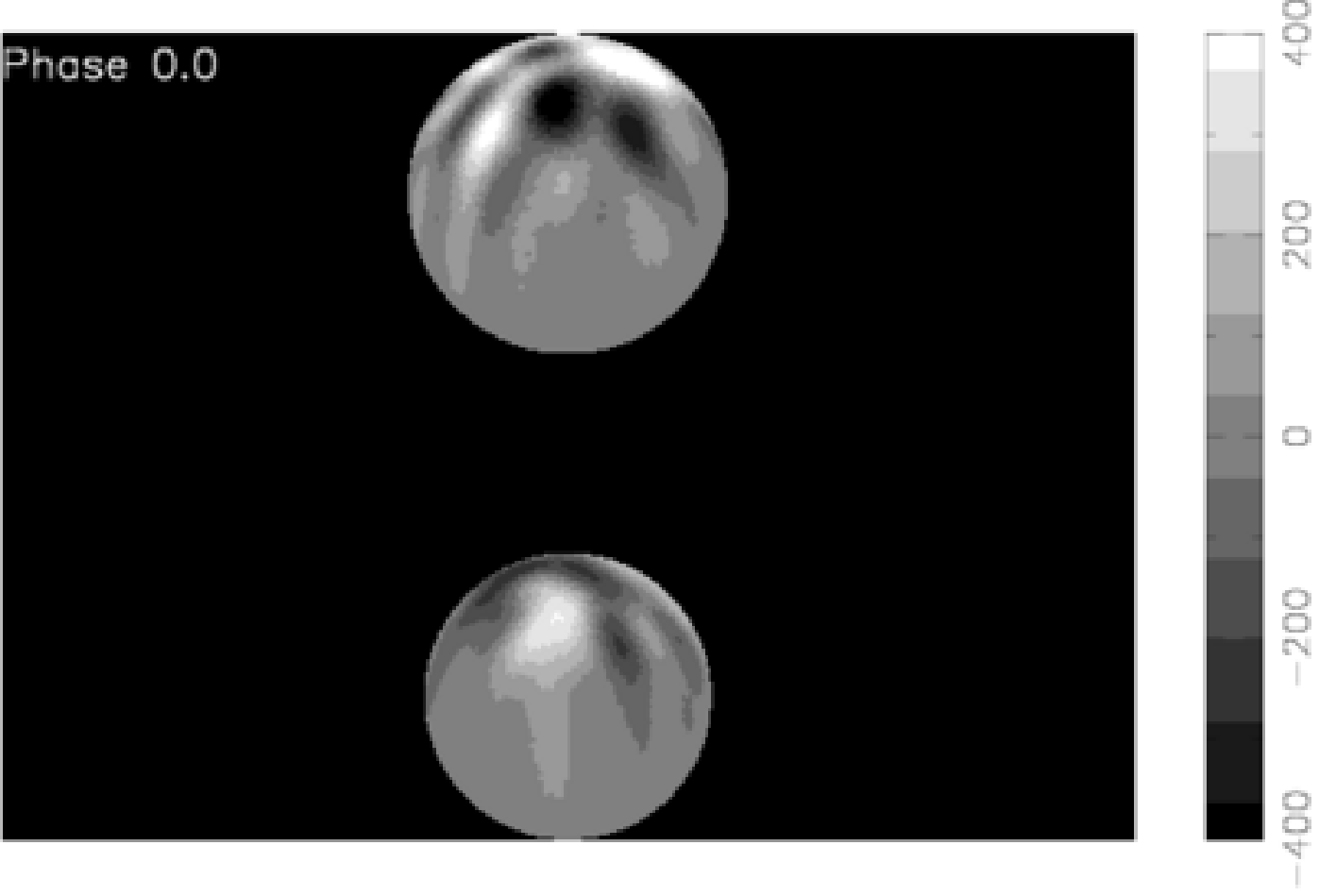} &
    \includegraphics[width=6.cm,angle=270]{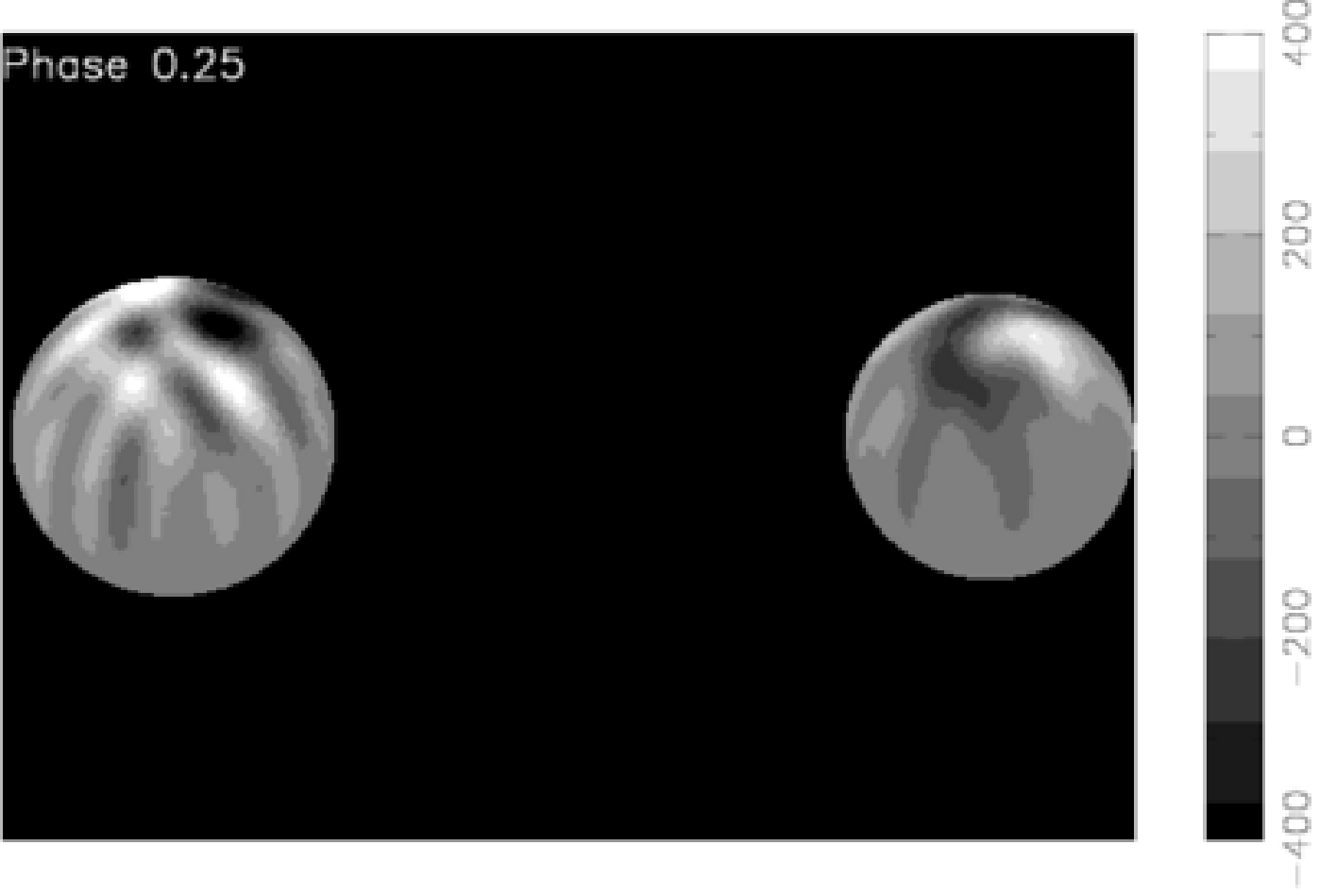} \\
    \includegraphics[width=6.cm,angle=270]{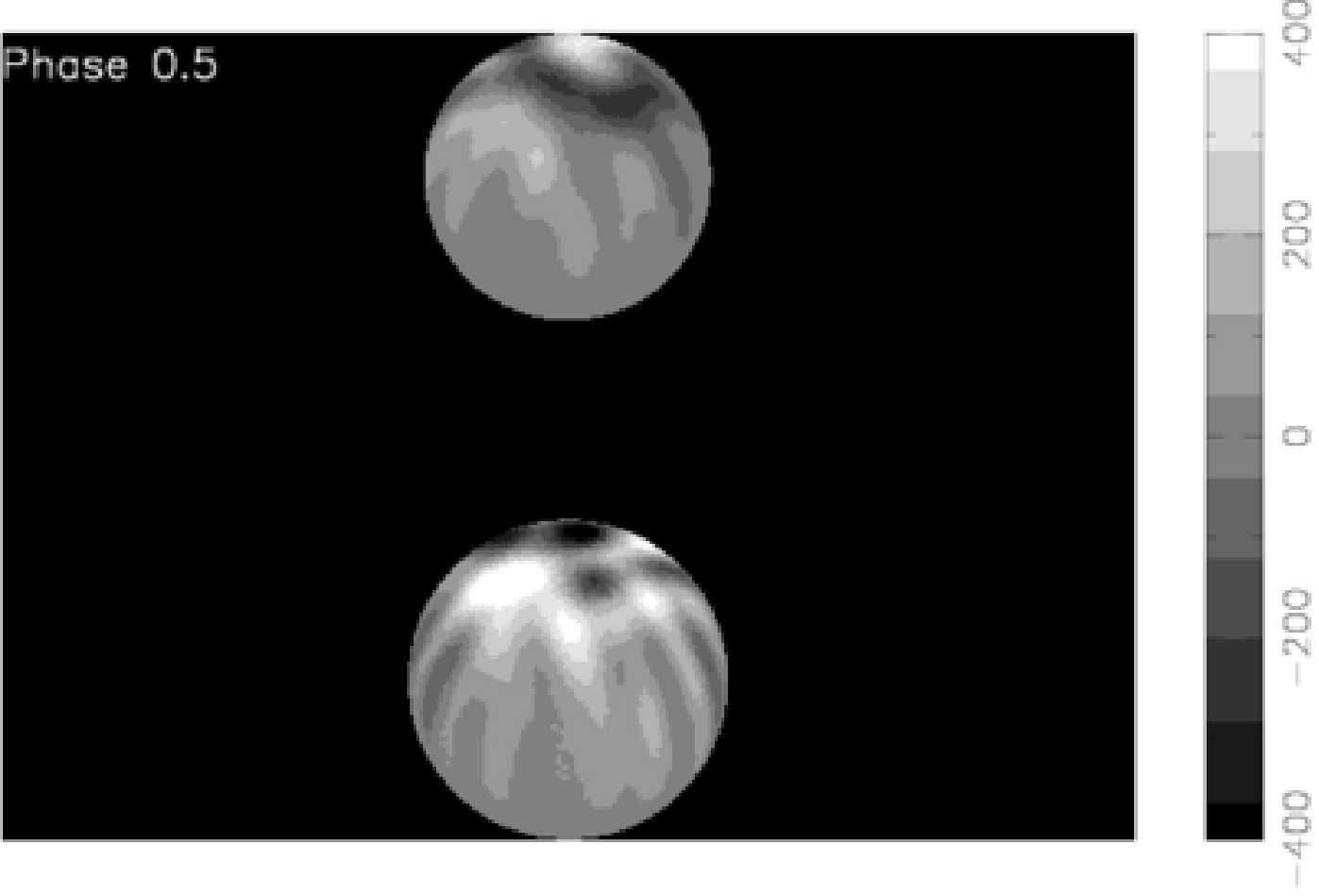} &
    \includegraphics[width=6.cm,angle=270]{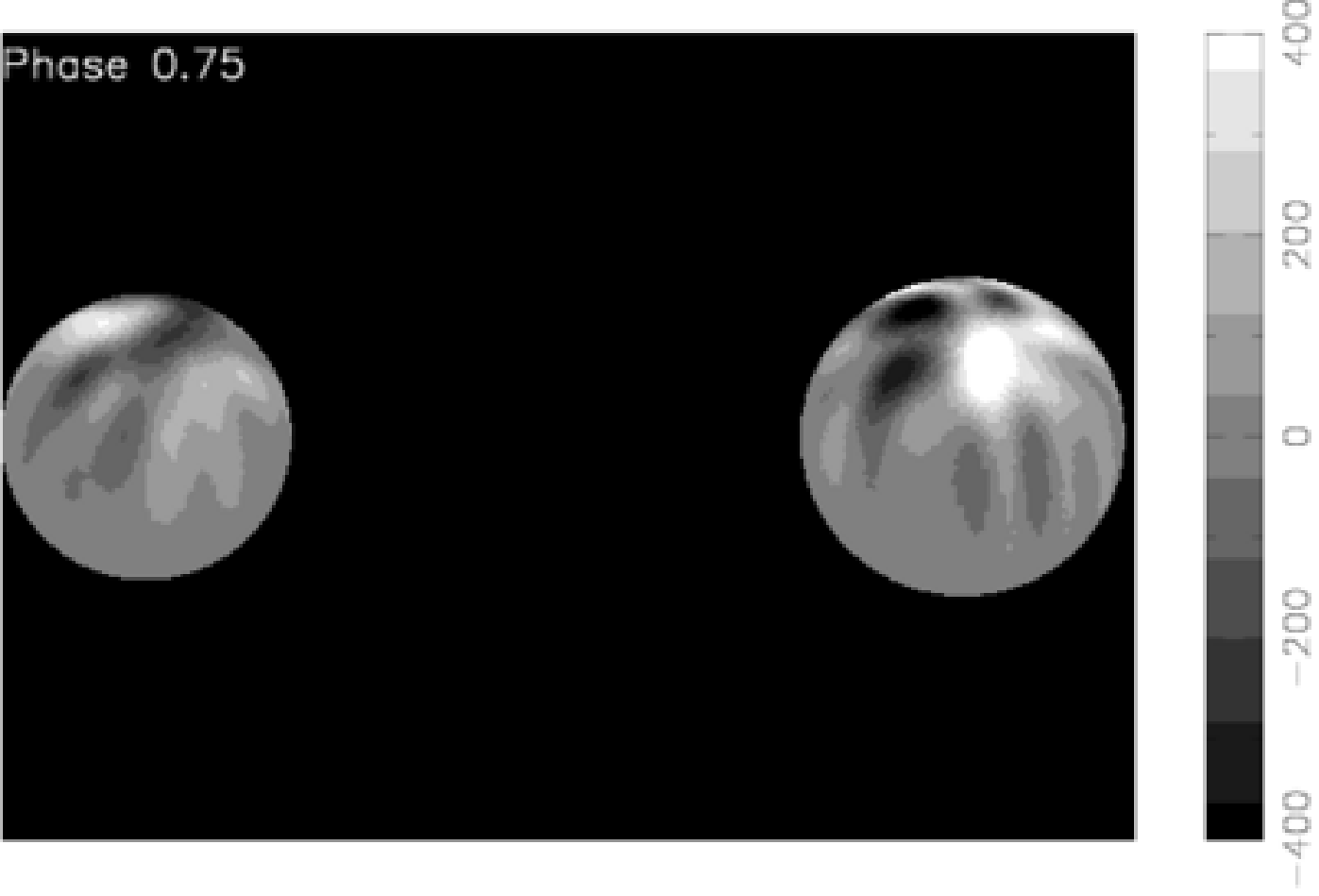} \\
  \end{tabular}
 \end{center}
\caption[Magnetic maps painted onto stellar surfaces]{The 2007 radial magnetic maps, in Fig. \ref{fig:magmaps}, are shown projected onto the surfaces of each star at the four quadrature phases. Note that we use a greyscale contrast range of $\pm400$ G which is a compromise of the different scales used for each star in Fig. \ref{fig:magmaps}.}
\protect\label{fig:magmovie}
\end{figure*}

In Fig. \ref{fig:magmovie} we show the radial magnetic field painted on to the surface of each star. This illustrates the field structure near the polar regions better than the mercator projection (Fig. \ref{fig:magmaps}). It also allows us to consider the distribution of flux on the star with respect to the binary orbital motion. We discuss the magnetic maps further in \S \ref{sect:discmag}, where we will also compare them to those obtained in 2004.

\section{2004 observations}
\protect\label{sect:2004}

A small number of observations were also acquired at the AAT of HD 155555 during 2004 September 23 - 28. The instrumental set-up and data reduction procedures were identical to those described in \S \ref{sect:obs}. The exposure time of each individual exposure was 300 s (compared to the 200 s of the 2007 observations). All orbital parameters of HD 155555 were re-calculated for the 2004 observations using the DoTS code and the \chisq\ minimisation technique (see \S \ref{sect:syspar}). The mass ratio ($q$) and semi-major velocity amplitude ($K_{1}$) were found to be consistent with the 2007 values (within the 2004 values uncertainties), therefore we adopted the higher precision 2007 values shown in Table \ref{tab:syspar}. The phase offset ($\phi_{0}$) and the radial velocity ($\gamma$) were also  re-calculated for the 2004 dataset. A phase offset of $\phi_{0}=0.752606$ (cf. $\phi_{0}=0.752474$, Table \ref{tab:syspar}) and a radial velocity of $\gamma=3.77$ \kms\ (cf. $\gamma=3.72\pm0.02$ \kms, Table \ref{tab:syspar}) were found. We note that the value found by \cite{strass00} for 1996 was considerably larger, $\gamma=5.9\pm0.2$ \kms. One possible source for changes in the apparent radial velocity of the HD 155555 is the dMe companion, LDS587B. However, a $\simeq$2 \kms\ change is probably too large to be attributed to the well separated LDS587B. Future observations will be required to determine if long-term trends exist in both $\phi_{0}$ and $\gamma$.

\begin{figure}
 \begin{center}
  \includegraphics[width=8cm,angle=270]{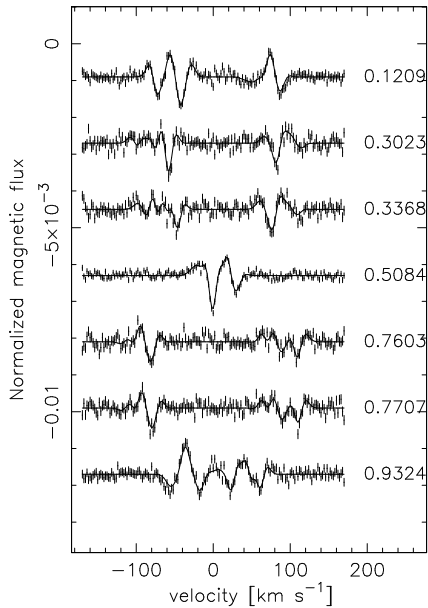}
 \end{center}
\caption[2004 Stoke V fits]{The observed 2004 Stokes V profiles from the LSD procedure are plotted along with their uncertainties. The solid line is the maximum entropy regularised fit we obtain to the data from Zeeman Doppler imaging. Profiles are ordered by orbital phase (listed to the right of each profile).}
\protect\label{fig:fitmag2004}
\end{figure}

Over the six nights a total of seven Stokes V spectra were taken. Two of these were taken close to each other on the same night and on the sixth night a very similar phase as was observed on the first night was obtained. Therefore together the Stokes V spectra effectively only sample five unique phases. The largest gap in the phase coverage is $\Delta\phi \simeq 0.25$ between phases $\phi$ = 0.51 - 0.76. The signal-to-noise in the peak order ranged from 180 - 220 on six of the nights and was 290 on September 25. In Fig. \ref{fig:fitmag2004} we show the 2004 Stokes V spectra and the fits. From these alone it is possible to ascertain that the field structure has changed on both stars. The secondary star generally shows a greater peak-to-peak amplitude that it did in 2007 and is now comparable to the primary star.

The relatively small number of observations in 2004 compared with the 2007 dataset result in lower quality maps with less detail and increased phase smearing. This is especially true for the brightness maps shown in Fig. \ref{fig:magmaps2004}. Here the weak low latitude spots are only recovered by pushing the maximum entropy regularised fits to the point where noise starts to appear. Despite this fact the 2004 data can still be compared with the 2007 data to provide valuable insight into the evolution over the intervening 2.5 year period, this is discussed in \S \ref{sect:discspot}.

The magnetic maps we recover are also shown in Fig. \ref{fig:magmaps2004}. The meridional magnetic maps show considerable cross-talk with the radial map. This much more severe in 2004 than the 2007 maps (Fig. \ref{fig:magmaps}). This is to be expected due to the poor phase coverage at this epoch (see \citealt{donati97recon} for further details). We note that on close comparison of the 2004 maps with those in Fig. \ref{fig:magmaps} (2007) it is apparent that regions in the 2004 maps are more stretched out in latitude than those in 2007. We suspect that this is a result of the poor phase sampling of the 2004 data. In order to assess the impact this has on the 2004 magnetic maps, we created a subset of seven spectra from the 2007 dataset that were nearest in phase to those of 2004. We then proceeded to use only these spectra in the ZDI process to re-produce the 2007 maps. The results of this procedure were that essentially the same maps were recovered however with increased latitudinal smearing present. This also led to a reduction by nearly 50\% in the field strengths of the recovered magnetic regions.

These results are not surprising given that the Doppler imaging process is inherently more sensitive to longitude than latitude position and it is only by repeatedly observing the same features as they move through the stellar rotational profile that we can accurately determine their latitude position. Also, as mentioned earlier, ZDI is only able to ascertain the orientation of the field by observing how the Stokes V signature behaves as it rotates into and out of our line of sight. Therefore the poorer the phase sampling the greater the cross-talk between the radial, azimuthal and meridional fields. Interestingly though, because this affects both stars equally, the ratio of the amplitudes of the field strengths recovered between the primary and secondary stars remained similar for this subset of 2007 data as for the whole dataset. We postpone further discussion of the 2004 magnetic maps to \S \ref{sect:discmag}.

\section{Coronal field extrapolations}
\protect\label{sect:corextrap}

The surface radial magnetic map of each star can be extrapolated to provide us with an initial idea of the coronal field structure using the coronal X-ray modelling technique of \cite{jardine02}, based around a code originally developed by \cite{vanball98}. At present we treat each star separately, making no attempt to model the interaction between the two stars that may occur. While this obviously places limitations on the conclusions we will be able to draw from the field extrapolations, it will have only minimal effect on either the small scale structure near to the stellar surface or the global orientation of the magnetic axis. This technique uses the ``Potential Field Source Surface'' method which assumes two boundary conditions, the first being that the radial field is equal to that recovered on the radial field map at the stellar surface and secondly that the field becomes purely radial at the source surface. The implication of this is that the magnetic field is forced open by the outward pressure of coronal gas at the source surface radius.

The further we get from the surface of a star the simpler the field geometry becomes. Eventually the outermost groups of closed field lines represent the lowest order, dipolar, field. From these we can establish whether or not the magnetic axis is aligned with the stellar rotational axis. The largely unseen lower hemisphere contains little magnetic flux. {{As discussed by \cite{jardine02structure}, the nature of the field in the unseen hemisphere can influence the coronal geometry of the low latitude field although the effect is much less apparent for the high-latitude field. The presence of  a binary companion may also change the connectivity of the largest-scale coronal field lines, since these may possibly connect the two stars. The orientation of the magnetic axis of the field, however, is not affected by either of these factors since it is determined purely by the surface field of the observed hemisphere.}} In order to illustrate the coronal topology we choose the source surface to be at a large enough radius for each star that the higher-order components of the field have died away and only the simplest, dipolar, structure remains. This is at a radius of 3.4 $\mathrm{R_*}$ for the primary and 6.8 $\mathrm{R_*}$ for the secondary component. The true locations of the source surface, and so the radii of the last closed field lines, will of course depend upon the density of X-ray emitting coronal gas. {{The value obtained for HD 155555 of $\log{(n_{e})}=10.7\pm0.23$ [cm$^{-3}$] determined by \cite{ness04} would suggest that the true source surface is probably smaller, perhaps comparable to the values determined for AB Dor by \cite{hussain07}.}} We note however that on single rapidly rotating stars stellar prominences have been observed at several stellar radii above the surface. For example, \cite{dunstone06} found that Speedy Mic had a complex prominence system located at twice the height above the stellar surface as the calculated co-rotation radius. This leads to the requirement of large closed magnetic loops in order to enforce co-rotation. In comparison the co-rotation radius of HD 155555 is in effect the orbital separation of 7.5 \rsun\ (by virtue of the fact that this is a tidally locked system).

In Fig. \ref{fig:corextrap} we show a representative sample of field lines from both regions close to the surface of the star and the larger scale field. As we described in the last section the radial surface maps of both stars, but the primary star in particular, show a complex magnetic field with many small magnetic regions. We can thus expect there to be significant structure at relatively small heights from the surface of each star. Such regions will produce small closed loops. This indeed is what we see from the coronal field extrapolations in Fig. \ref{fig:corextrap}.

\section{Discussion}
\protect\label{sect:disc}

\subsection{Distribution of stellar spots}
\protect\label{sect:discspot}

In addition to the September 2004 and April 2007 brightness maps presented in this work, there are two previously published Doppler images of HD 155555. \cite{hatzes99} present observations taken in September 1990 and \cite{strass00} those obtained in May 1996. These four well spaced epochs provide a valuable opportunity to study the evolution and stability of the spot groups in this tidally locked binary system. The details of the different Doppler imaging codes, the phase coverage and signal-to-noise of observations mean that we have to be careful in our interpretation of the strength and sizes of individual features that we see. However a general comparison of the distribution of spots should be possible.

Considering the primary star first, all four maps show a polar spot. In 1990 \cite{hatzes99} observed a significantly decentred polar spot that was tilted in the direction of the secondary star. The other three epochs show a more centrally located polar cap that extends down to a latitude of +60\degr. The well sampled 2007 dataset shown in Figs. \ref{fig:spotmaps} and \ref{fig:spotmovie} suggest that the polar cap is not homogeneous but is composed of a number of large dark spots. All four epochs show a lack of mid-latitude spots and a band of low latitude spots at approximately +30\degr. This bi-modality in the latitudinal distribution of spots seems to be a consistent feature of the G5IV primary. This is reminiscent of the brightness maps found for G dwarfs in the young \aper\ cluster by \cite{barnes98} and more recently of the G dwarf HD 171488 by \cite{marsden06}. All these stars show spots emerging near the polar or at low latitudes but with a distinct lack of spots at intermediate latitudes. {{Theoretical flux-tube modelling by \cite{granzer00} predict a bimodal spot distribution for very young pre-main sequence stars. \cite{strass00} considered these models for the particular case of HD 155555 and found that the predictions did not match the observed spot patterns. Instead, meridional flows may explain how flux is transported to the poles. Or else two distinct flux emergence mechanisms may be in operation.}}

\cite{hatzes99} revealed a string of such low-latitude spots on the anti-facing hemisphere (i.e. the side of the primary facing {\it{away}} from the secondary star). This was also reported by \cite{strass00}, although their maps recover spots covering a larger range of stellar longitudes and therefore also revealed a number of spots on the trailing hemisphere (the one that always follows behind with respect to the binary orbital motion). Both our maps for 2004 and 2007 reveal a band of low-latitude spots. The more reliable 2007 maps show that this covers three quarters of the stellar circumference and shows a preference for spots to be located on the facing and trailing hemispheres. The gap that breaks what would otherwise be a complete ring of low-latitude spots occurs between phases $\phi$ = 0.2 - 0.5. This corresponds to the leading/facing hemisphere of the primary star and is thus in total contrast to the 1990 map. We are therefore left to conclude that there are no apparent stable `active longitudes' on the primary star of the HD 155555 system.

\begin{figure*}
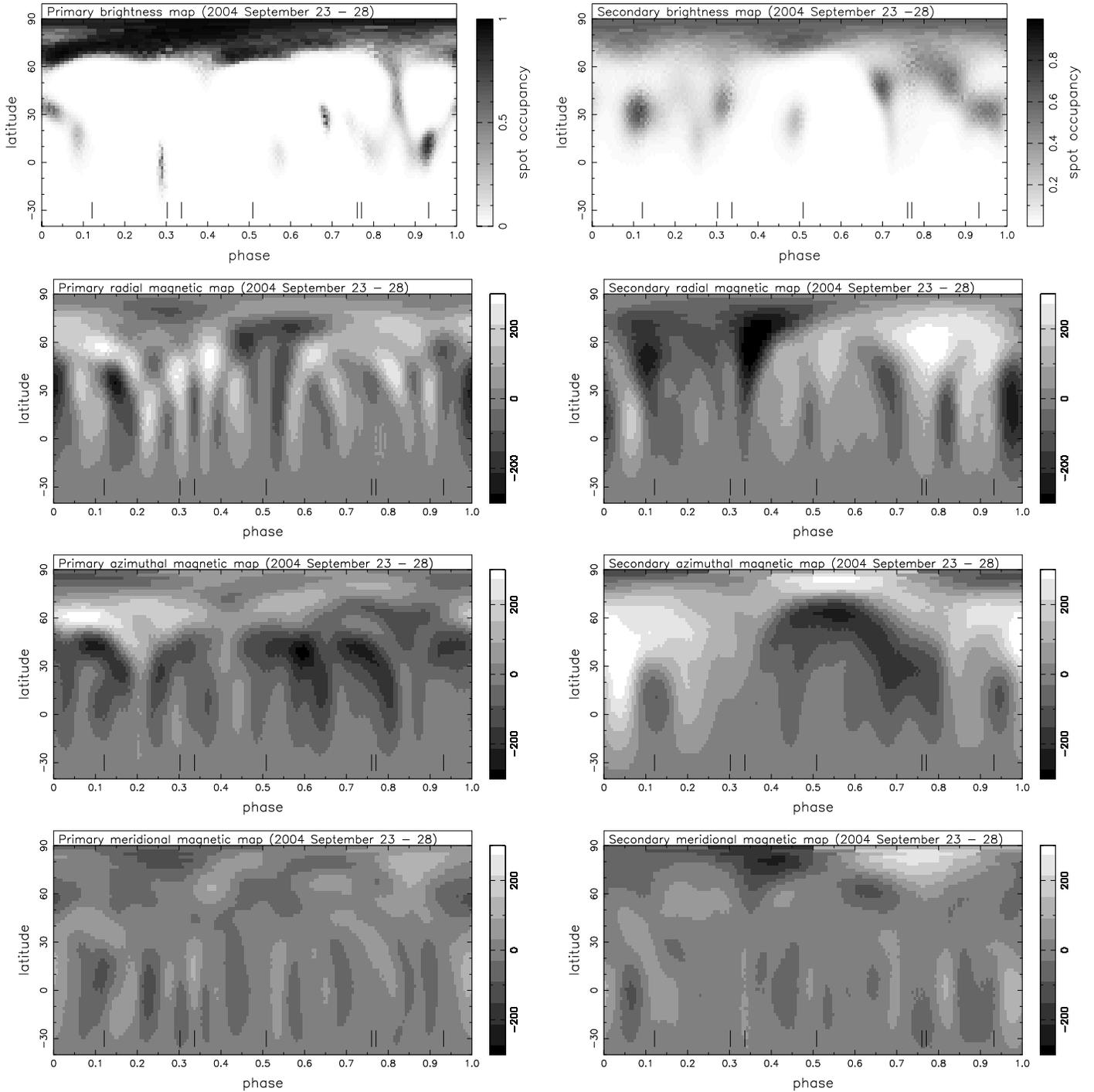

 \begin{center}
  \begin{tabular}{cc}
    \includegraphics[width=4.5cm,angle=270]{Image/njd_fig21.ps} &
    \includegraphics[width=4.5cm,angle=270]{Image/njd_fig22.ps} \\
    \includegraphics[width=4.5cm,angle=270]{Image/njd_fig23.ps} &
    \includegraphics[width=4.5cm,angle=270]{Image/njd_fig24.ps} \\
    \includegraphics[width=4.5cm,angle=270]{Image/njd_fig25.ps} &
    \includegraphics[width=4.5cm,angle=270]{Image/njd_fig26.ps} \\
    \includegraphics[width=4.5cm,angle=270]{Image/njd_fig27.ps} &
    \includegraphics[width=4.5cm,angle=270]{Image/njd_fig28.ps} \\
  \end{tabular}
\caption[2004 Maps]{Brightness (top) and magnetic maps (middle - radial field, bottom - azimuthal field) are shown for September 2004 observations. Note that unlike 2007 maps, the greyscale contrast range of the magnetic maps is the same between the primary and secondary maps, as illustrated in the key to the immediate right of each map.}
\protect\label{fig:magmaps2004}
\end{center}
\end{figure*}

With the exception of 2004, the secondary star does not have a true polar cap but instead has a high-latitude spot that just touches the pole. \cite{hatzes99} found such a spot tilted towards the primary star (at phase $\phi\simeq0.55$) in 1990, whereas \cite{strass00} recovered a similar feature but on the trailing hemisphere (at phase $\phi\simeq0.30$) of the secondary star in 1996. Our 2004 spot map shows the secondary star to possess a true (centrally located) polar spot. In 2007 we again find a high-latitude spot that this time is centred on phase $\phi\simeq0.4$  and so is again tilted in the direction of the primary star. Given our relatively poor understanding of how polar spots form, it is possible that the shifting location of the polar spot may be linked to the extreme tilt of the magnetic axis found in \S \ref{sect:corextrap}.

{{All epochs reveal spots at mid and low-latitudes on the secondary star. In particular a large low-latitude (30\degr) spot group is seen at {\it{all}} epochs between phases $\phi=0.1 - 0.3$ (trailing hemisphere). As this appears to be a consistent feature of the secondary star maps it may indeed indicate an active spot longitude.}}

\subsection{Surface magnetic topology and dynamo processes}
\protect\label{sect:discmag}

\begin{figure*}
 \begin{center}
  \begin{tabular}{cc}
    \includegraphics[width=7.cm,angle=0]{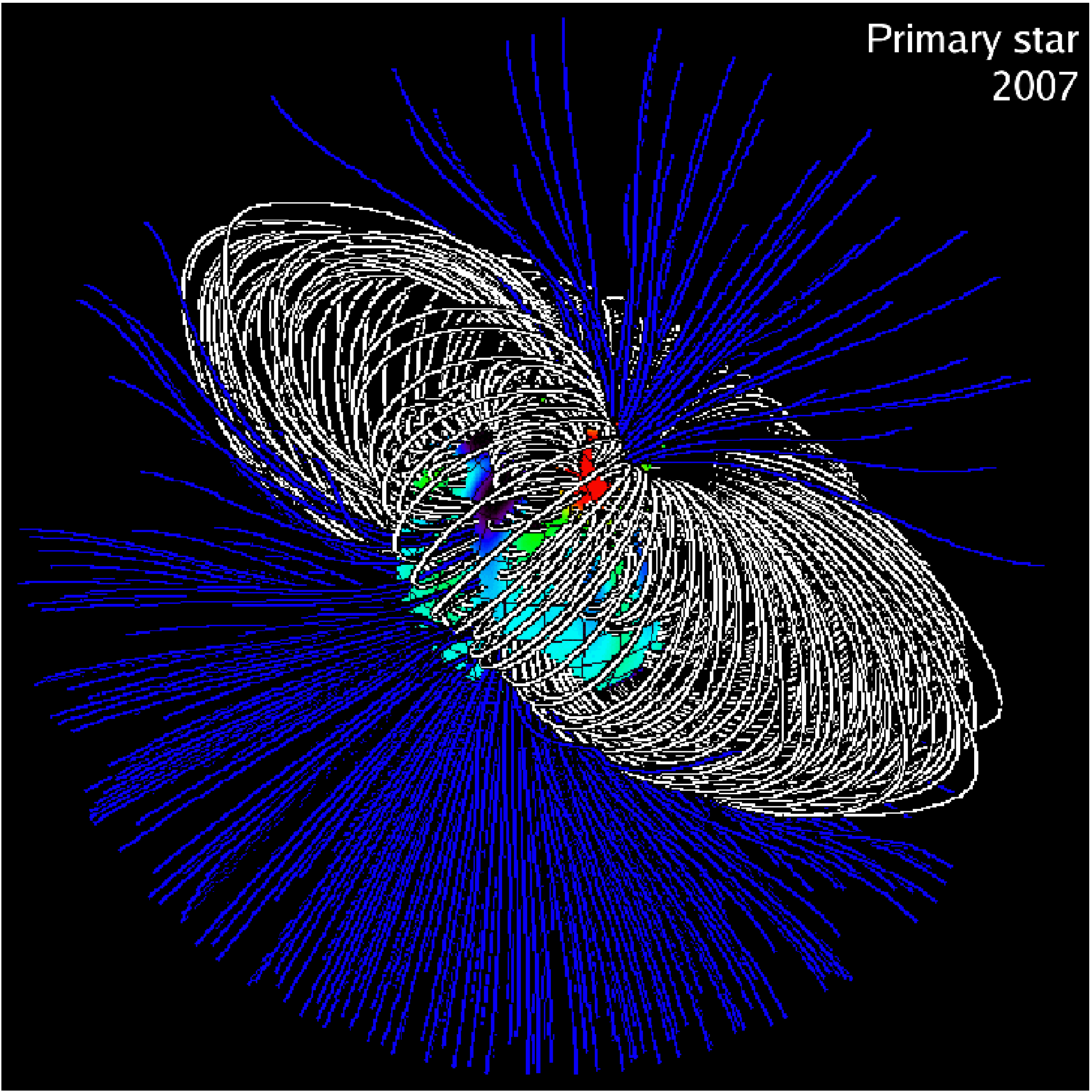} &
    \includegraphics[width=7.cm,angle=0]{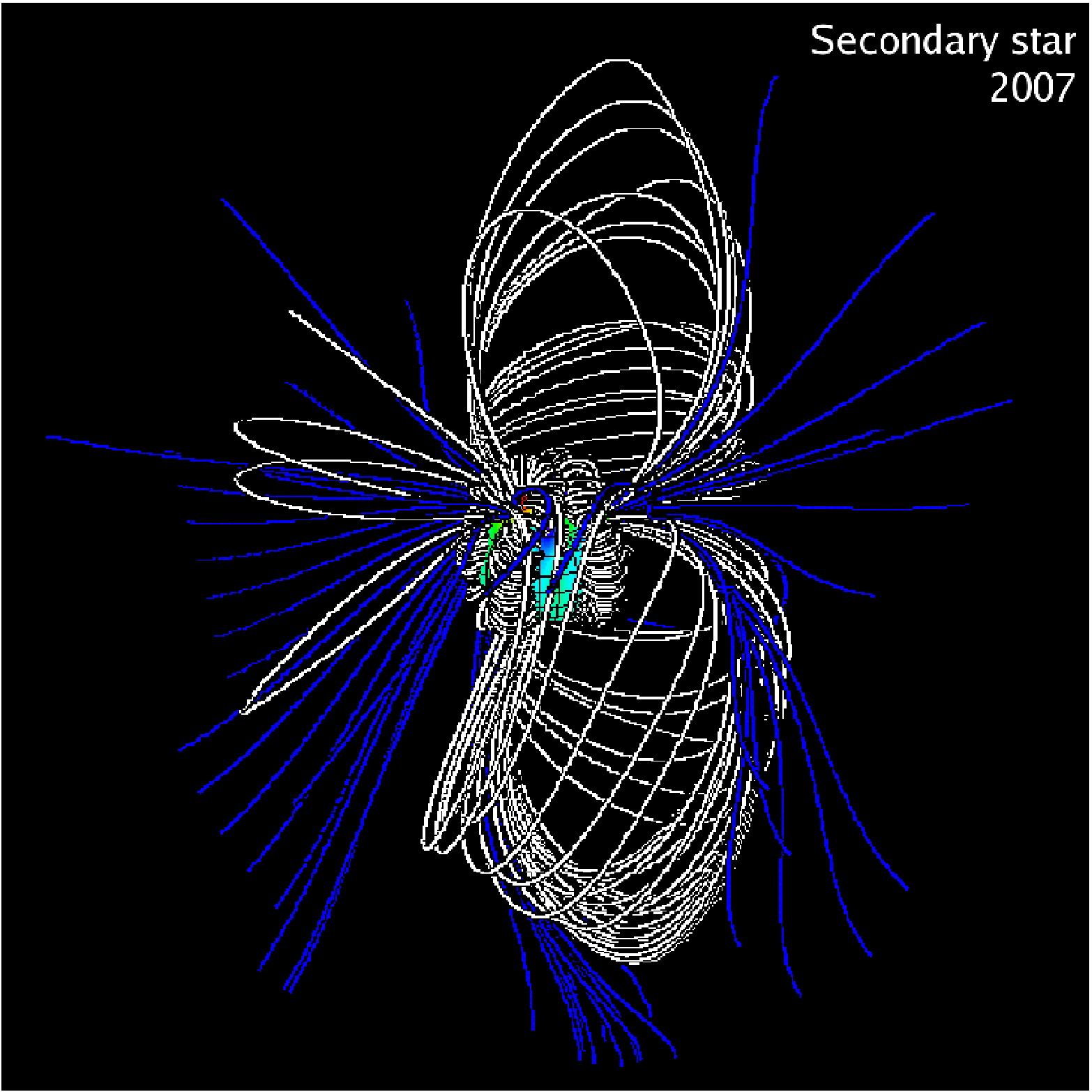} \\
    \includegraphics[width=7.cm,angle=0]{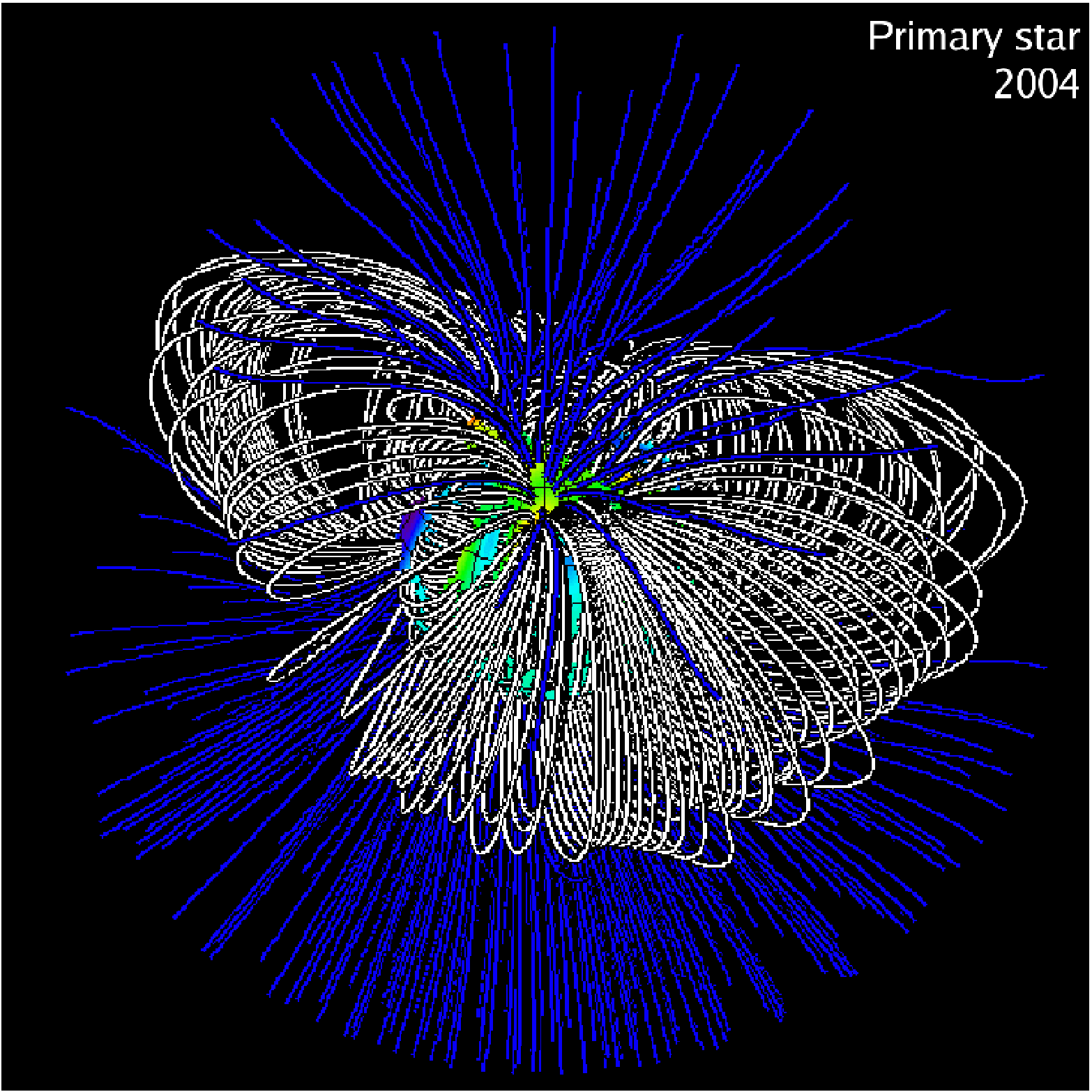} &
    \includegraphics[width=7.cm,angle=0]{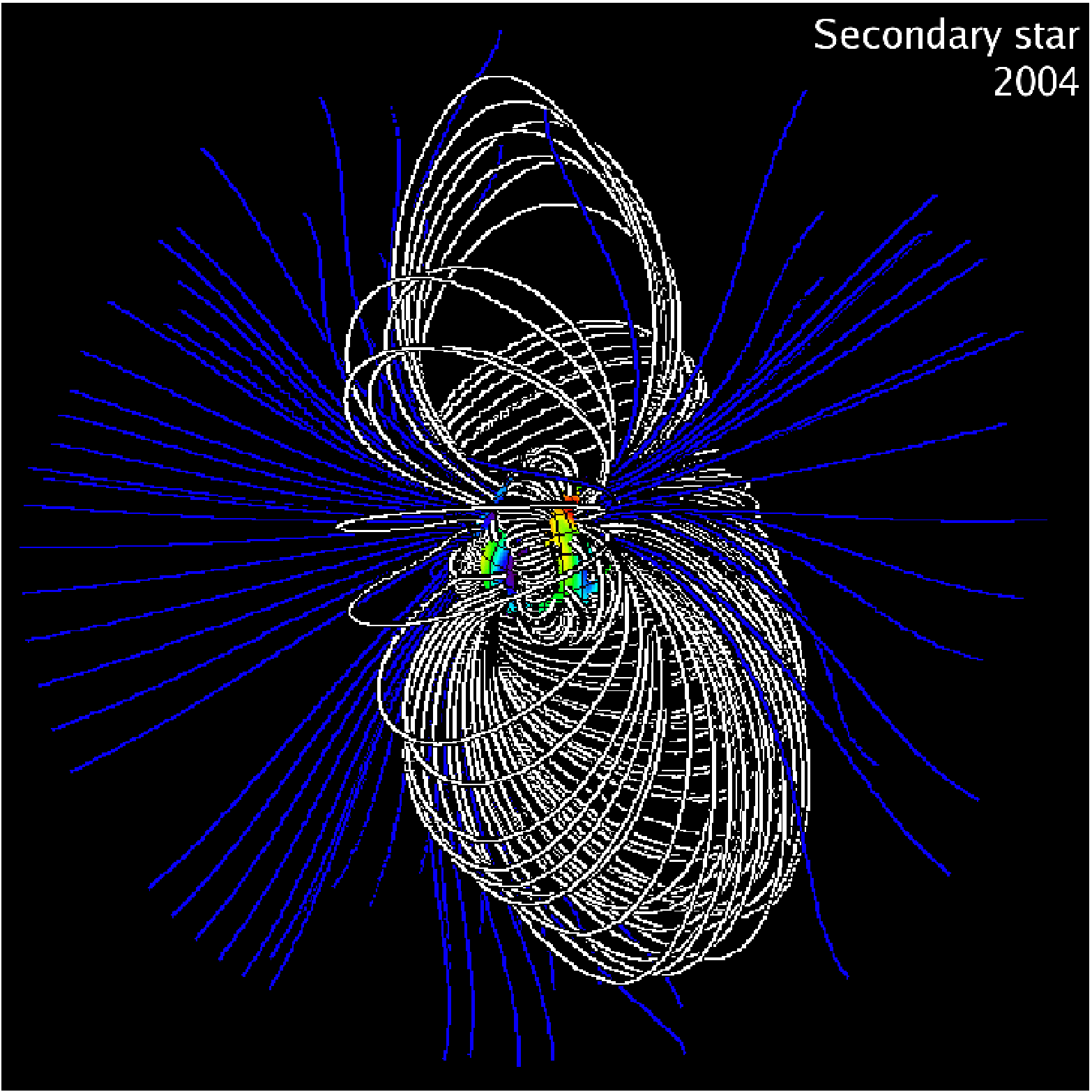} \\
   \end{tabular}
 \end{center}
\caption[Coronal field extrapolations]{Coronal field extrapolations are shown for the primary star (left panels) and secondary star (right panels) for 2007 data (top panels) and 2004 data (bottom panels). These particular phases were chosen as they best illustrate the large scale field orientation. The primary is shown at phase $\phi=0.8$ (and so near to the trailing hemisphere), while the secondary is at phase $\phi=0.9$ (nearest to the anti-facing hemisphere). {{Open and closed field lines are represented by blue and white lines respectively. The radial magnetic maps are shown on the stellar surfaces with blue corresponding to negative, green to zero and red to positive magnetic flux.}}}
\protect\label{fig:corextrap}
\end{figure*}

The magnetic maps we produced in the last sections (Figs. \ref{fig:magmaps} and \ref{fig:magmaps2004}) are the first of a pre-main sequence binary system.  Both components of HD 155555 possess large regions of strong azimuthal field near to the stellar surface. These are of equivalent strength (in fact, often stronger than) the recovered radial field. Strong azimuthal field is recovered on all active cool stars that have been mapped so far using ZDI. This is in contrast to the relatively small amounts of horizontal field found on the comparatively inactive Sun. Solar azimuthal field is thought to be buried deep in the base of the convective envelope where dynamo activity is generated. Previous authors (e.g. \citealt{donati03} and \citealt{petit04}) have therefore postulated that the presence of strong azimuthal field, at the stellar surface of these rapidly rotating stars, suggests that their dynamo may be active throughout the convection zone. While this suggestion is contrary to the classical understanding of the generation of large-scale fields in slowly rotating (solar-type) stars, it has recently gained support from the numerical simulations of \cite{brown07}. These authors find that global-scale toroidal and poloidal magnetic fields can be built and maintained in the bulk of the convection zone of rapidly rotating stars, despite the presence of turbulent convection. In particular \cite{brown07} find strong ordered torodial fields which may explain the observation, as in HD 155555, of strong surface azimuthal field. We can now compare the maps of both components of HD 155555 to each other and look for evolution between the 2004 and 2007 epochs.

\subsubsection{Primary star}

\begin{figure}
 \begin{center}
  \begin{tabular}{c}
    \includegraphics[width=8.cm,angle=0]{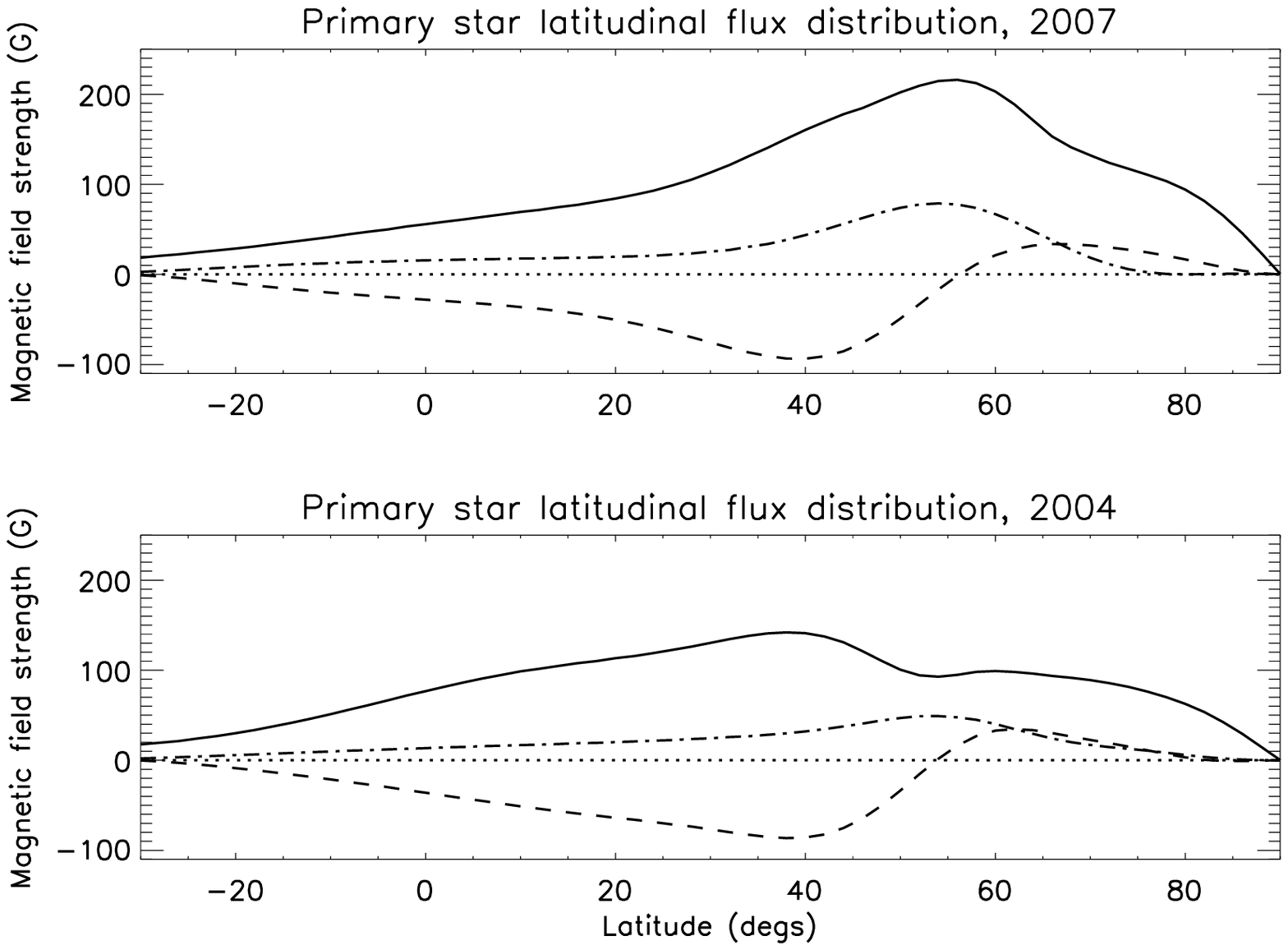} \\ 	              \includegraphics[width=8.cm,angle=0]{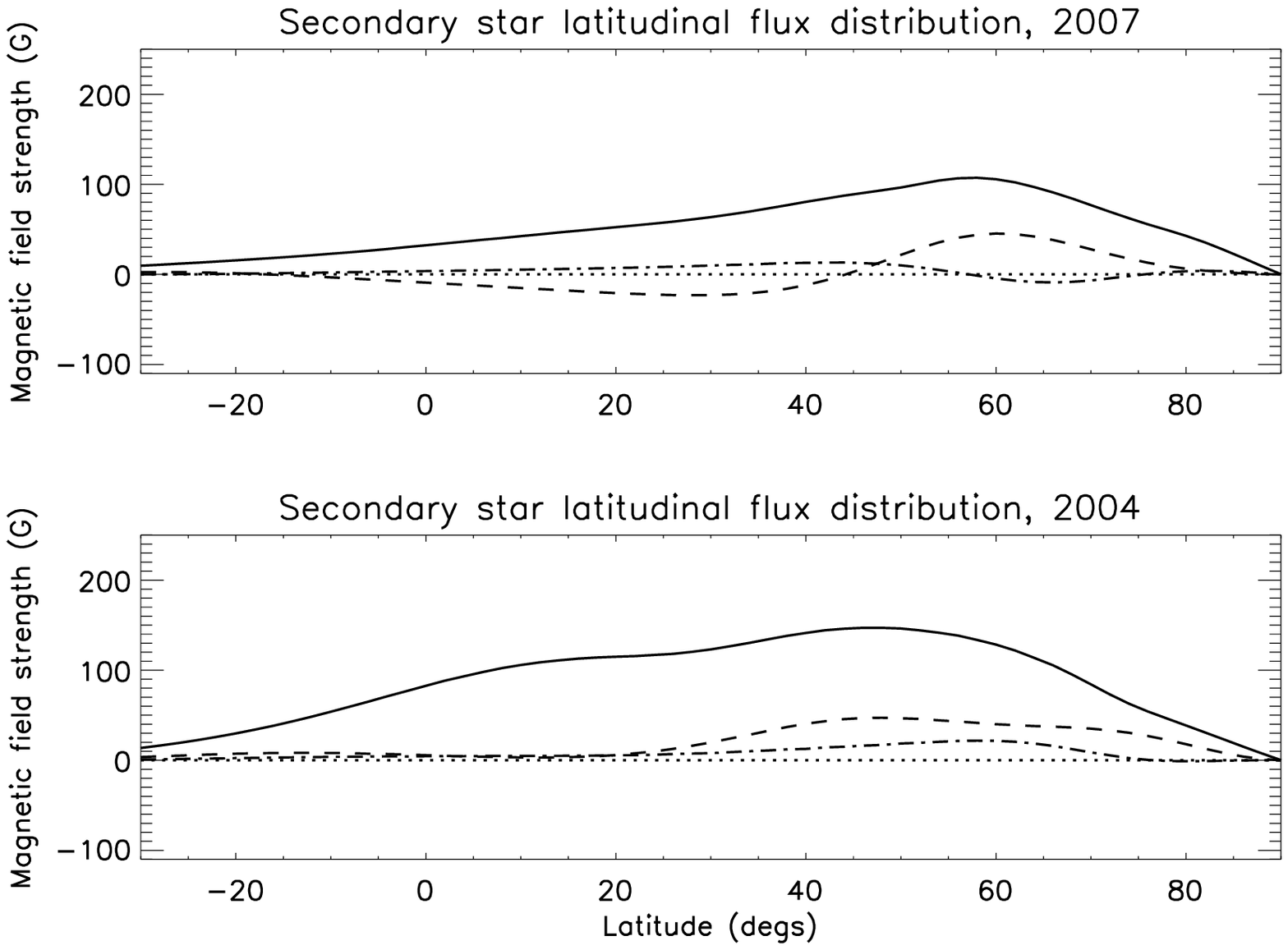}
  \end{tabular}
 \end{center}
\caption[Latitudinal flux distribution]{The latitudinal flux distribution is shown for each star on both the 2004 and 2007 epochs. In each panel the solid line represents the quadratic magnetic flux, the dashed line shows the azimuthal field and the dash-dot line depicts the radial field. The value of a given quantity is the integration over all longitudes and has been corrected for the relative pixel area at that particular latitude - hence the reduction in all fluxes observed at the pole. }
\protect\label{fig:latdist}
\end{figure}

The radial field map of the primary star shows many small regions of flux of alternating polarities. A visual inspection of the 2007 primary radial maps reveals no real latitude dependence, with both positive and negative field at all latitudes and no preference at the pole. In 2004 the maps do show a slight preference for positive polarities, especially at mid to high latitudes (60 - 80\degr). The radial field maps are integrated over all longitudes in Fig. \ref{fig:latdist} to test for the presence of a latitudinal polarity dominance. In doing so considerable cancellation of the flux occurs due to the already described complex nature of the maps. However, Fig. \ref{fig:latdist} shows that we are left with a preference for positive radial flux at all latitudes and for both epochs, with a peak flux at 55\degr.

The azimuthal field maps of the primary star reveal a considerably simpler field structure than the radial maps. This is particularly apparent in 2007 where the surface topology can essentially be described by just two regions. A large region of positive field is seen encircling the star at high latitudes, centred on ($\simeq$75\degr). This points down to lower latitudes ($\simeq$60\degr) at phase $\phi=0.8$. It therefore resembles a ring of positive flux that is tilted by 15\degr\ with respect to the rotational axis. Due to the fact that the ring crosses over the rotational pole, part of it then appears as a negative polarity region. The other region is of negative field at lower latitudes. In 2004 the azimuthal maps appear more complex that in 2007, with more mixing of positive and negative polarities. The general latitude distribution is similar to 2007, with positive field at high latitudes and negative field at lower latitudes. However, the field appears more axisymmetric in 2004 with both rings of flux being better centred on the rotation pole. Again, in Fig. \ref{fig:latdist} we integrate the azimuthal maps over all longitudes to find that the primary star indeed shows a dominant negative polarity at all latitudes below 55\degr, but then switches to positive field at high latitudes. The similarity of the azimuthal flux distribution in both 2004 and 2007 suggests that this is a relatively stable feature of the primary star.

The meridional field maps show two very high latitude regions of opposite polarity in both 2007 and 2004. This is exactly what you would expect to result from a homogeneously oriented ring of horizontal field that crosses the pole. The ZDI process will recover part of the field as azimuthal (as discussed above) and part as meridional. The low latitude features (as already outlined in \S \ref{sect:magmaps} and \S \ref{sect:2004}) are the result of cross-talk from the radial field (particularly severe in 2004). Therefore, the azimuthal and high-latitude meridional flux appears to be attributable to the toroidal field, while the radial and low-latitude meridional flux is part of the poloidal field.

\subsubsection{Secondary star}

The secondary star has a considerably simpler radial field map than the primary star. Like the corresponding map of the primary star, the radial field map appears to have no latitude dependence on the polarity of the field. Indeed when we collapse the maps in longitude in Fig. \ref{fig:latdist} we achieve virtually total cancellation of the positive and negative fluxes. In 2007 the residual trend is for positive polarity at latitudes lower than 55\degr and negative flux at higher latitudes. In 2004 there is a slight indication of preference for positive polarity for both field components. 

The maps of azimuthal field of the secondary star are unusually complex. In 2007 we recover low latitude flux of mixed polarity, with a complex region composed of both negative and positive field at phases $\phi\simeq0.6-0.9$. However, we see a positive ring of field at high latitudes, again tilted (like the azimuthal field of the primary star in 2007) away from the rotational axis by $\simeq15$\degr\ and towards phase $\phi=0.3$. Fig. \ref{fig:latdist} shows that negative azimuthal field dominates at latitudes below 45\degr\ then the field becomes positive at higher latitudes. Similarly in 2004, a ring of non-axisymmetric positive azimuthal field can be seen at high-latitudes. Fig. \ref{fig:latdist} shows a preference for positive azimuthal field at all latitudes. Both epochs have a strong, low latitude, region of negative azimuthal field centred on phase $\phi\simeq0.5$ (the hemisphere facing the primary star). The meridional maps of the secondary star show two high-latitude regions of opposite polarity. They are therefore similar to those of the primary star and we thus refer the reader to the discussion of the primary star's meridional maps in the last section.

\subsection{Magnetic regions, polar spots and axis misalignment }
\protect\label{sect:discaxis}

{{To summarise the last section; we find evidence for rings of azimuthal field which are non-axisymmetric but yet show a clear constant polarity structure. The trend for both stars of HD 155555 is to have a high-latitude ring of positive azimuthal field which is tilted with respect to the rotational axis and then for this to be underlined by negative field at low latitudes. Interestingly the rings of positive azimuthal field closely follow the polar spots discussed in \S \ref{sect:discspot}. \cite{donati99lqhya_hr1099} and \cite{donati03} found a similar pattern between the azimuthal field and polar spots on HR 1099.}}

{{The similarity of the 2004 and 2007 coronal extrapolations shown in Fig. \ref{fig:corextrap} is striking.  On both epochs the secondary star exhibits an extreme tilt of its magnetic axis. From plotting the neutral polarity line at the source surface we calculate that the magnetic axis is approximately 75\degs\ misaligned with the rotational axis.  The primary stars magnetic field is more closely aligned to that of its rotational axis. We find a $\simeq$30\degs\ misalignment to that of the rotational axis in 2004 and a larger $\simeq$45\degs\ misalignment in 2007. This may be related to the fact that, as discussed in the last section, the ring of surface azimuthal field shows a tilt by approximately 15\degr\ for the primary star in 2007 but more of an axisymmetric ring in 2004. So it appears that both the radial and azimuthal fields were both more axisymmetric in 2004. It therefore seems likely that the surface azimuthal field is linked to the orientation of the large scale field.}}

\subsection{Relative field strengths and magnetic energy}

In the 2007 maps shown in Figs. \ref{fig:magmaps} and \ref{fig:magmaps2004} the magnetic field strength scale needed to be different for each star and yet for 2004 they could be the same. The average quadratic field intensity can be calculated for both stars at each epoch. This can also be examined graphically, as a function of latitude, from the solid line in Fig. \ref{fig:latdist}. We find for the well sampled 2007 epoch an average field strength (integrated field modulus over the whole stellar surface) of 68 G for the primary star and 43 G for the secondary star. So in 2007 the primary star appeared to have a field strength 50 \% greater than that of the secondary star. For the 2004 maps we find a field strength of 52 G for the primary star and 61 G for the secondary. It therefore appears that the field strength of the secondary star was considerably stronger in 2004, both in absolute terms (42\% stronger than in 2007) and relative to that of the primary star.

As mentioned in \S \ref{sect:2004} the comparatively poor phase sampling of 2004 data must be taken into account. We found that using only a subset of the 2007 data (choosing the same number of spectra available in 2004) resulted in an almost 50\% reduction in the recovered flux. In addition the observed flux was spread out in stellar latitude. The effect of this can be seen in Fig. \ref{fig:latdist} where the total quadratic flux (solid line) shows a broader distribution in the 2004 panels than those of 2007. The effect of fewer phases of observation is that we effectively miss (do not recover) flux from the stellar surface. Therefore the 16 G weaker field strength of the primary star in 2004 is likely due, at least in part, to this effect. However, reduced phase sampling certainly cannot explain why the secondary star has a field strength 18 G (42\%) {\it{stronger }} in 2004 than in 2007. In fact it suggests quite to the contrary, the real difference in the flux between 2004 and 2007 must have been even greater. 

Another quantity which we can calculate in an attempt to understand the above observations is that of the magnetic energy. This can be calculated for each of the radial and azimuthal fields. We can then quote the energy stored in each field component as a fraction of the total energy. The meridional field is found to contain between 8 - 12 \% of the total magnetic energy (depending upon the star and epoch). It is therefore small compared to the radial and azimuthal contributions and we do not discuss it further. We find that in 2007 the azimuthal field of the primary star stores 58\% of the total energy with 34\% in the radial component. This is similar to the case of AB Dor and LQ Hya where the azimuthal field is found to preferentially store more of the field energy than the radial field at most epochs (\citealt{donati03}). In the case of the secondary star, in 2007 the radial field dominates by storing 48\% of the magnetic energy while the azimuthal field has 43\%. 

When we perform the same calculation for the 2004 maps we find that the azimuthal and radial fields store equal energy (44\%) on the primary star, corresponding to a decrease in the relative azimuthal field strength. However on the secondary star the opposite appears to have occurred, the azimuthal field is found to contain 54\% of the energy and the radial field with 38\%. We again note that the poor sampling of the 2004 dataset may have an impact on these results. Observations with very poor phase sampling can be prone to cross-talk between the radial and azimuthal field components. This would lead to a reduction in the contrast between the two field components and may partly explain the observation of more equal percentages for the primary star. Also if, by chance, a particularly strong region of field (be it azimuthal or radial) was missed by our observations then this would be reflected in the balance of the magnetic energy. The situation would be considerably worse in the case of poor phase {\it{coverage}}. If the observations were very poorly spaced then there would be little constraint on the flux for large parts of the stellar surface. While the 2004 observations are poorly sampled they are at least reasonably well spaced. We therefore conclude that the relative strengths of the field components for 2004 should be treated with caution and as an indication only.

In summary, it appears that the magnetic field of the secondary star has evolved considerably between 2004 and 2007. The average field strength has become much weaker (both in absolute value and relative to the primary star) and the azimuthal field component appears to have weakened relative to that of the radial. This is accompanied by an apparent change in the azimuthal topology from a comparatively simple field (characterised by a ring of positive flux) in 2004 to a more complex distribution in 2007. If, as suggested by \cite{donati03}, the strong azimuthal field observed on active stars is closely related to the toroidal component of the large-scale dynamo field then the observed behaviour may represent the decay phase of an activity cycle. There is insufficient evidence to make a similar comment about the primary star at this time.

\subsection{Comparing HD 155555 with other stars}
\protect\label{sect:compstar}
The HD 155555 system is particularly interesting as it is a young binary system and is therefore a potentially very promising target for disentangling the effects of youth and binarity. As a close binary system, the components of HD 155555 are tidally locked just like the primary stars in evolved RS CVn binaries, such as HR 1099. However, with an age of around 18 Myr (see \S \ref{sect:atmo}) the system is still a pre-main sequence object and therefore younger than the well studied zero-age main-sequence rapidly rotating single stars, AB Dor and LQ Hya (50 - 100 Myr, e.g. \citealt{janson07}). This fact is reflected in the bloated size of both components of HD 155555. Despite the fact that AB Dor and LQ Hya are of approximately equal mass to HD 155555 we find, in \S \ref{sect:syspar}, radii of 1.5 and 1.3 \rsun\ for the primary and secondary star respectively (c.f. $\approx$ 1 \rsun\ for AB Dor and LQ Hya, \citealt{donati03b}). This will lead to increased convection zone depths of both components of HD 155555 relative to AB Dor and LQ Hya. Therefore, somewhat ironically, both the evolved HR 1099 binary (R=3.7 \rsun) and HD 155555 have convection zone depths greater than AB Dor and LQ Hya. Consequently if convection zone depth uniquely determines the observed magnetic surface topologies, then our attempt to compare the effects of binarity and youth will be ultimately inconclusive.

As already mentioned, the observation of strong surface azimuthal field is common to HD 155555, AB Dor, LQ Hya and HR 1099. Beyond this, HR 1099 is characterised by easily identified rings of azimuthal field that are normally axisymmetric. A similar ring of azimuthal field is often observed on AB Dor but only at high-latitudes. We find that both components of HD 155555 show rings of azimuthal field, however they are often non-axisymmetric. HR 1099 also possesses a latitudinal dependence on the polarity of the radial field with a unipolar polar cap. In contrast the young rapid rotators AB Dor and LQ Hya show mixed polarities of radial field at all latitudes, including the pole. At first glance this is essentially what is seen for HD 155555 but when the latitudinal distribution of flux is plotted a clear trend is seen. It can be seen in Fig. \ref{fig:latdist} that the primary star (in both 2004 and 2007) has a bias towards positive radial flux at all latitudes but increasing so towards high-latitudes. \cite{petit04hr1099} confirmed the presence of an anti-correlation of the latitudinal distribution of azimuthal and radial field on HR 1099. This is not seen on AB Dor or LQ Hya but is present on the Sun. Fig. \ref{fig:latdist} shows tentative evidence of a similar anti-correlation for HD 155555 in 2007 for the secondary star and for the primary star at latitudes below 55\degr\ in both 2007 and 2004. However, this does not hold for the secondary star in 2004 where both the radial and azimuthal field is positively orientated. Further observations of HD 155555 will be needed to establish long-term trends.

\subsection{The coronal fields of HD 155555 and binary interaction}
\protect\label{sect:disccor}

In \S \ref{sect:corextrap} we used the recovered radial field maps to extrapolate the coronal structure of both stars. Due to the small binary separation of 7.5 \rsun\ ($\simeq5.5\ R_{*}$, see Table \ref{tab:syspar}) there is no doubt that the magnetospheres of the two stars will interact. The question is: what will the nature of the interaction be? Due to the relative complexity of the surface radial field maps of both stars, the extrapolated coronal field is also correspondingly complex. As Fig. \ref{fig:corextrap} shows, at locations between the stars where the two fields would interact, both fields retain sufficient complexity that their interaction will not be simply modelled by that of two dipolar fields. 

It is clear that the next step will be to perform a full binary extrapolation, where the radial field maps of both stars are considered simultaneously. This will model which field lines will preferentially connect with the other star rather than to regions on the same star. The upper limit on the longevity of such `binary field lines' will largely be defined by the relative differential rotation rates of each star. The surface rotation properties of HD 155555 will be the subject of a forthcoming paper.

\section{Conclusions}
\protect\label{sect:conc}

We have produced the first simultaneous magnetic maps of both components of a binary system. These show that the technique of Zeeman Doppler imaging can be adapted to the binary case. This opens up the potential to study many short period pre-main sequence and main sequence binary systems where both components have Stokes V detections. The HD 155555 system provides an interesting first insight into the magnetic fields of a young binary system. In general, we find that the resulting magnetic maps appear to share features in common with both young rapidly rotating stars and the evolved primary stars of RS CVn binaries. The most surprising result is the severe misalignment of the secondary stars magnetic and rotational axes. The impact of this can be seen on the radial and azimuthal maps. Also interesting is the significant weakening of the secondary star's magnetic field strength from 2004 to 2007. Especially as this appears to be associated with a decline in the proportion of magnetic energy stored in the azimuthal field. 

Once a binary potential field extrapolation code has been completed it will be possible not only to more reliably model the coronal structure of both stars but also their interaction. This will allow us to predict the locations of stellar winds and determine which field lines are capable of containing hot X-ray emitting gas. We can therefore look forward to learning more about the mass loss rates from binary systems and modelling the X-ray properties of binaries, including flare loop lengths and energies.

\section{ACKNOWLEDGEMENTS}

We would like to thank the staff at the Anglo-Australian Telescope for their support. NJD acknowledges the financial support of a UK STFC studentship.  

\bibliographystyle{mn2e}
\bibliography{iau_journals,master,ownrefs,njd2}

\begin{thebibliography}{}

\bibitem[\protect\citeauthoryear{{Bagnuolo} Jr. \& {Gies}}{{Bagnuolo} \&
  {Gies}}{1991}]{bagnuolo91}
{Bagnuolo} Jr. W.~G.,  {Gies} D.~R.,  1991, ApJ, 376, 266

\bibitem[\protect\citeauthoryear{{Barnes}, {Collier Cameron}, {James} \&
  {Donati}}{{Barnes} et~al.}{2000}]{barnes00}
{Barnes} J.~R.,  {Collier Cameron} A.,  {James} D.~J.,    {Donati} J.-F.,
  2000, MNRAS, 314, 162

\bibitem[\protect\citeauthoryear{{Barnes}, {Collier Cameron}, {Unruh}, {Donati}
  \& {Hussain}}{{Barnes} et~al.}{1998}]{barnes98}
{Barnes} J.~R.,  {Collier Cameron} A.,  {Unruh} Y.~C.,  {Donati} J.~F.,
  {Hussain} G.~A.~J.,  1998, MNRAS, 299, 904

\bibitem[\protect\citeauthoryear{{Barnes}, {Lister}, {Hilditch} \& {Collier
  Cameron}}{{Barnes} et~al.}{2004}]{barnes04}
{Barnes} J.~R.,  {Lister} T.~A.,  {Hilditch} R.~W.,    {Collier Cameron} A.,
  2004, MNRAS, 348, 1321

\bibitem[\protect\citeauthoryear{{Bennett}, {Evans} \& {Laing}}{{Bennett}
  et~al.}{1967}]{bennett67}
{Bennett} N.~W.~W.,  {Evans} D.~S.,    {Laing} J.~D.,  1967, MNRAS, 137, 107

\bibitem[\protect\citeauthoryear{{Brown}, {Browning}, {Brun}, {Miesch},
  {Nelson} \& {Toomre}}{{Brown} et~al.}{2007}]{brown07}
{Brown} B.~P.,  {Browning} M.~K.,  {Brun} A.~S.,  {Miesch} M.~S.,  {Nelson}
  N.~J.,    {Toomre} J.,  2007, in American Institute of Physics Conference
  Series Vol.~948 of American Institute of Physics Conference Series, {Strong
  Dynamo Action in Rapidly Rotating Suns}.
pp 271--278

\bibitem[\protect\citeauthoryear{{Carlsson}, {Rutten}, {Bruls} \&
  {Shchukina}}{{Carlsson} et~al.}{1994}]{carlsson94}
{Carlsson} M.,  {Rutten} R.~J.,  {Bruls} J.~H.~M.~J.,    {Shchukina} N.~G.,
  1994, A\&A, 288, 860

\bibitem[\protect\citeauthoryear{{Collier Cameron}}{{Collier
  Cameron}}{1997}]{cam97dots}
{Collier Cameron} A.,  1997, MNRAS, 287, 556

\bibitem[\protect\citeauthoryear{{Collier-Cameron} \&
  {Unruh}}{{Collier-Cameron} \& {Unruh}}{1994}]{cam94twotemp}
{Collier-Cameron} A.,  {Unruh} Y.~C.,  1994, MNRAS, 269, 814

\bibitem[\protect\citeauthoryear{{D'Antona} \& {Mazzitelli}}{{D'Antona} \&
  {Mazzitelli}}{1997}]{d.antona97}
{D'Antona} F.,  {Mazzitelli} I.,  1997, Memorie della Societa Astronomica
  Italiana, 68, 807

\bibitem[\protect\citeauthoryear{{Dempsey}, {Linsky}, {Fleming} \&
  {Schmitt}}{{Dempsey} et~al.}{1993}]{dempsey93}
{Dempsey} R.~C.,  {Linsky} J.~L.,  {Fleming} T.~A.,    {Schmitt} J.~H.~M.~M.,
  1993, ApJS, 86, 599

\bibitem[\protect\citeauthoryear{{Dempsey}, {Neff} \& {Lim}}{{Dempsey}
  et~al.}{2001}]{dempsey01}
{Dempsey} R.~C.,  {Neff} J.~E.,    {Lim} J.,  2001, AJ, 122, 332

\bibitem[\protect\citeauthoryear{{Donati}, {Jardine}, {Petit}, {Morin},
  {Bouvier}, {Cameron}, {Delfosse}, {Dintrans}, {Dobler}, {Dougados},
  {Ferreira}, {Forveille}, {Gregory}, {Harries}, {Hussain}, {Menard} \&
  {Paletou}}{{Donati} et~al.}{2007}]{donati07}
{Donati} J.~.,  {Jardine} M.~M.,  {Petit} P.,  {Morin} J.,  {Bouvier} J.,
  {Cameron} A.~C.,  {Delfosse} X.,  {Dintrans} B.,  {Dobler} W.,  {Dougados}
  C.,  {Ferreira} J.,  {Forveille} T.,  {Gregory} S.~G.,  {Harries} T.,
  {Hussain} G.~A.~J.,  {Menard} F.,    {Paletou} F.,  2007, ArXiv Astrophysics
  e-prints

\bibitem[\protect\citeauthoryear{{Donati}}{{Donati}}{1999}]{donati99lqhya_hr10%
99}
{Donati} J.-F.,  1999, MNRAS, 302, 457

\bibitem[\protect\citeauthoryear{{Donati} \& {Brown}}{{Donati} \&
  {Brown}}{1997}]{donati97recon}
{Donati} J.-F.,  {Brown} S.~F.,  1997, A\&A, 326, 1135

\bibitem[\protect\citeauthoryear{{Donati}, {Cameron}, {Semel}, {Hussain},
  {Petit}, {Carter}, {Marsden}, {Mengel}, {L{\'o}pez Ariste}, {Jeffers} \&
  {Rees}}{{Donati} et~al.}{2003}]{donati03}
{Donati} J.-F.,  {Cameron} A.~C.,  {Semel} M.,  {Hussain} G.~A.~J.,  {Petit}
  P.,  {Carter} B.~D.,  {Marsden} S.~C.,  {Mengel} M.,  {L{\'o}pez Ariste} A.,
  {Jeffers} S.~V.,    {Rees} D.~E.,  2003, MNRAS, 345, 1145

\bibitem[\protect\citeauthoryear{{Donati} \& {Collier Cameron}}{{Donati} \&
  {Collier Cameron}}{1997}]{donati97ab}
{Donati} J.-F.,  {Collier Cameron} A.,  1997, MNRAS, 291, 1

\bibitem[\protect\citeauthoryear{{Donati}, {Collier Cameron}, {Hussain} \&
  {Semel}}{{Donati} et~al.}{1999}]{donati99ab}
{Donati} J.-F.,  {Collier Cameron} A.,  {Hussain} G.~A.~J.,    {Semel} M.,
  1999, MNRAS, 302, 437

\bibitem[\protect\citeauthoryear{{Donati}, {Collier Cameron} \&
  {Petit}}{{Donati} et~al.}{2003}]{donati03b}
{Donati} J.-F.,  {Collier Cameron} A.,    {Petit} P.,  2003, MNRAS, 345, 1187

\bibitem[\protect\citeauthoryear{{Donati}, {Semel}, {Carter}, {Rees} \&
  {Collier Cameron}}{{Donati} et~al.}{1997}]{donati97survey}
{Donati} J.-F.,  {Semel} M.,  {Carter} B.~D.,  {Rees} D.~E.,    {Collier
  Cameron} A.,  1997, MNRAS, 291, 658

\bibitem[\protect\citeauthoryear{Dunstone}{Dunstone}{2008}]{dunstone08thesis}
Dunstone N.,  2008, PhD thesis, University of St Andrews, St Andrews, Scotland
  (in preparation)

\bibitem[\protect\citeauthoryear{{Dunstone}, {Barnes}, {Cameron} \&
  {Jardine}}{{Dunstone} et~al.}{2006}]{dunstone06}
{Dunstone} N.~J.,  {Barnes} J.~R.,  {Cameron} A.~C.,    {Jardine} M.,  2006,
  MNRAS, 365, 530

\bibitem[\protect\citeauthoryear{{Granzer}, {Sch{\"u}ssler}, {Caligari} \&
  {Strassmeier}}{{Granzer} et~al.}{2000}]{granzer00}
{Granzer} T.,  {Sch{\"u}ssler} M.,  {Caligari} P.,    {Strassmeier} K.~G.,
  2000, A\&A, 355, 1087

\bibitem[\protect\citeauthoryear{{Hatzes} \& {K{\"u}rster}}{{Hatzes} \&
  {K{\"u}rster}}{1999}]{hatzes99}
{Hatzes} A.~P.,  {K{\"u}rster} M.,  1999, A\&A, 346, 432

\bibitem[\protect\citeauthoryear{{Hussain}, {Donati}, {Collier Cameron} \&
  {Barnes}}{{Hussain} et~al.}{2000}]{hussain00}
{Hussain} G.~A.~J.,  {Donati} J.-F.,  {Collier Cameron} A.,    {Barnes} J.~R.,
  2000, MNRAS, 318, 961

\bibitem[\protect\citeauthoryear{{Hussain}, {Jardine}, {Donati}, {Brickhouse},
  {Dunstone}, {Wood}, {Dupree}, {Collier Cameron} \& {Favata}}{{Hussain}
  et~al.}{2007}]{hussain07}
{Hussain} G.~A.~J.,  {Jardine} M.,  {Donati} J.-F.,  {Brickhouse} N.~S.,
  {Dunstone} N.~J.,  {Wood} K.,  {Dupree} A.~K.,  {Collier Cameron} A.,
  {Favata} F.,  2007, MNRAS, 377, 1488

\bibitem[\protect\citeauthoryear{{Janson}, {Brandner}, {Lenzen}, {Close},
  {Nielsen}, {Hartung}, {Henning} \& {Bouy}}{{Janson} et~al.}{2007}]{janson07}
{Janson} M.,  {Brandner} W.,  {Lenzen} R.,  {Close} L.,  {Nielsen} E.,
  {Hartung} M.,  {Henning} T.,    {Bouy} H.,  2007, A\&A, 462, 615

\bibitem[\protect\citeauthoryear{{Jardine}, {Collier Cameron} \&
  {Donati}}{{Jardine} et~al.}{2002}]{jardine02structure}
{Jardine} M.,  {Collier Cameron} A.,    {Donati} J.-F.,  2002, MNRAS, 333, 339

\bibitem[\protect\citeauthoryear{{Jardine}, {Wood}, {Collier Cameron}, {Donati}
  \& {Mackay}}{{Jardine} et~al.}{2002}]{jardine02}
{Jardine} M.,  {Wood} K.,  {Collier Cameron} A.,  {Donati} J.-F.,    {Mackay}
  D.~H.,  2002, MNRAS, 336, 1364

\bibitem[\protect\citeauthoryear{{Kupka}, {Piskunov}, {Ryabchikova}, {Stempels}
  \& {Weiss}}{{Kupka} et~al.}{1999}]{vald99}
{Kupka} F.,  {Piskunov} N.,  {Ryabchikova} T.~A.,  {Stempels} H.~C.,    {Weiss}
  W.~W.,  1999, A\&AS, 138, 119

\bibitem[\protect\citeauthoryear{{Kurucz}}{{Kurucz}}{1993}]{kurucz93}
{Kurucz} R.,  1993, ATLAS9 Stellar Atmosphere Programs and 2 km/s grid.~Kurucz
  CD-ROM No.~13.~ Cambridge, Mass.: Smithsonian Astrophysical Observatory,
  1993., 13

\bibitem[\protect\citeauthoryear{{Marsden}, {Donati}, {Semel}, {Petit} \&
  {Carter}}{{Marsden} et~al.}{2006}]{marsden06}
{Marsden} S.~C.,  {Donati} J.-F.,  {Semel} M.,  {Petit} P.,    {Carter} B.~D.,
  2006, MNRAS, 370, 468

\bibitem[\protect\citeauthoryear{{Ness}, {G{\"u}del}, {Schmitt}, {Audard} \&
  {Telleschi}}{{Ness} et~al.}{2004}]{ness04}
{Ness} J.-U.,  {G{\"u}del} M.,  {Schmitt} J.~H.~M.~M.,  {Audard} M.,
  {Telleschi} A.,  2004, A\&A, 427, 667

\bibitem[\protect\citeauthoryear{{Pasquini}, {Cutispoto}, {Gratton} \&
  {Mayor}}{{Pasquini} et~al.}{1991}]{pasquini91}
{Pasquini} L.,  {Cutispoto} G.,  {Gratton} R.,    {Mayor} M.,  1991, A\&A, 248,
  72

\bibitem[\protect\citeauthoryear{{Petit}, {Donati}, {Oliveira}, {Auri{\`e}re},
  {Bagnulo}, {Landstreet}, {Ligni{\`e}res}, {L{\"u}ftinger}, {Marsden},
  {Mouillet}, {Paletou}, {Strasser}, {Toqu{\'e}} \& {Wade}}{{Petit}
  et~al.}{2004}]{petit04}
{Petit} P.,  {Donati} J.-F.,  {Oliveira} J.~M.,  {Auri{\`e}re} M.,  {Bagnulo}
  S.,  {Landstreet} J.~D.,  {Ligni{\`e}res} F.,  {L{\"u}ftinger} T.,  {Marsden}
  S.,  {Mouillet} D.,  {Paletou} F.,  {Strasser} S.,  {Toqu{\'e}} N.,    {Wade}
  G.~A.,  2004, MNRAS, 351, 826

\bibitem[\protect\citeauthoryear{{Petit}, {Donati}, {Wade}, {Landstreet},
  {Bagnulo}, {L{\"u}ftinger}, {Sigut}, {Shorlin}, {Strasser}, {Auri{\`e}re} \&
  {Oliveira}}{{Petit} et~al.}{2004}]{petit04hr1099}
{Petit} P.,  {Donati} J.-F.,  {Wade} G.~A.,  {Landstreet} J.~D.,  {Bagnulo} S.,
   {L{\"u}ftinger} T.,  {Sigut} T.~A.~A.,  {Shorlin} S.~L.~S.,  {Strasser} S.,
  {Auri{\`e}re} M.,    {Oliveira} J.~M.,  2004, MNRAS, 348, 1175

\bibitem[\protect\citeauthoryear{{Piskunov}, {Kupka}, {Ryabchikova}, {Weiss} \&
  {Jeffery}}{{Piskunov} et~al.}{1995}]{vald95}
{Piskunov} N.~E.,  {Kupka} F.,  {Ryabchikova} T.~A.,  {Weiss} W.~W.,
  {Jeffery} C.~S.,  1995, A\&AS, 112, 525

\bibitem[\protect\citeauthoryear{{Semel}}{{Semel}}{1989}]{semel89}
{Semel} M.,  1989, A\&A, 225, 456

\bibitem[\protect\citeauthoryear{{Semel}, {Donati} \& {Rees}}{{Semel}
  et~al.}{1993}]{semel93}
{Semel} M.,  {Donati} J.-F.,    {Rees} D.~E.,  1993, A\&A, 278, 231

\bibitem[\protect\citeauthoryear{{Sestito} \& {Randich}}{{Sestito} \&
  {Randich}}{2005}]{sestito05}
{Sestito} P.,  {Randich} S.,  2005, A\&A, 442, 615

\bibitem[\protect\citeauthoryear{Skilling \& Bryan}{Skilling \&
  Bryan}{1984}]{skilling84}
Skilling J.,  Bryan R.~K.,  1984, MNRAS, 211, 111

\bibitem[\protect\citeauthoryear{{Strassmeier}, {Hall}, {Fekel} \&
  {Scheck}}{{Strassmeier} et~al.}{1993}]{strass93}
{Strassmeier} K.~G.,  {Hall} D.~S.,  {Fekel} F.~C.,    {Scheck} M.,  1993,
  A\&AS, 100, 173

\bibitem[\protect\citeauthoryear{{Strassmeier} \& {Rice}}{{Strassmeier} \&
  {Rice}}{2000}]{strass00}
{Strassmeier} K.~G.,  {Rice} J.~B.,  2000, A\&A, 360, 1019

\bibitem[\protect\citeauthoryear{{Unruh} \& {Collier Cameron}}{{Unruh} \&
  {Collier Cameron}}{1995}]{unruh95}
{Unruh} Y.~C.,  {Collier Cameron} A.,  1995, MNRAS, 273, 1

\bibitem[\protect\citeauthoryear{{Valenti} \& {Piskunov}}{{Valenti} \&
  {Piskunov}}{1996}]{valenti96}
{Valenti} J.~A.,  {Piskunov} N.,  1996, A\&AS, 118, 595

\bibitem[\protect\citeauthoryear{{van Ballegooijen}, {Cartledge} \&
  {Priest}}{{van Ballegooijen} et~al.}{1998}]{vanball98}
{van Ballegooijen} A.~A.,  {Cartledge} N.~P.,    {Priest} E.~R.,  1998, ApJ,
  501, 866

\end{thebibliography}

\newpage

\begin{figure*}
 \begin{center}
  \includegraphics[width=12cm,angle=270]{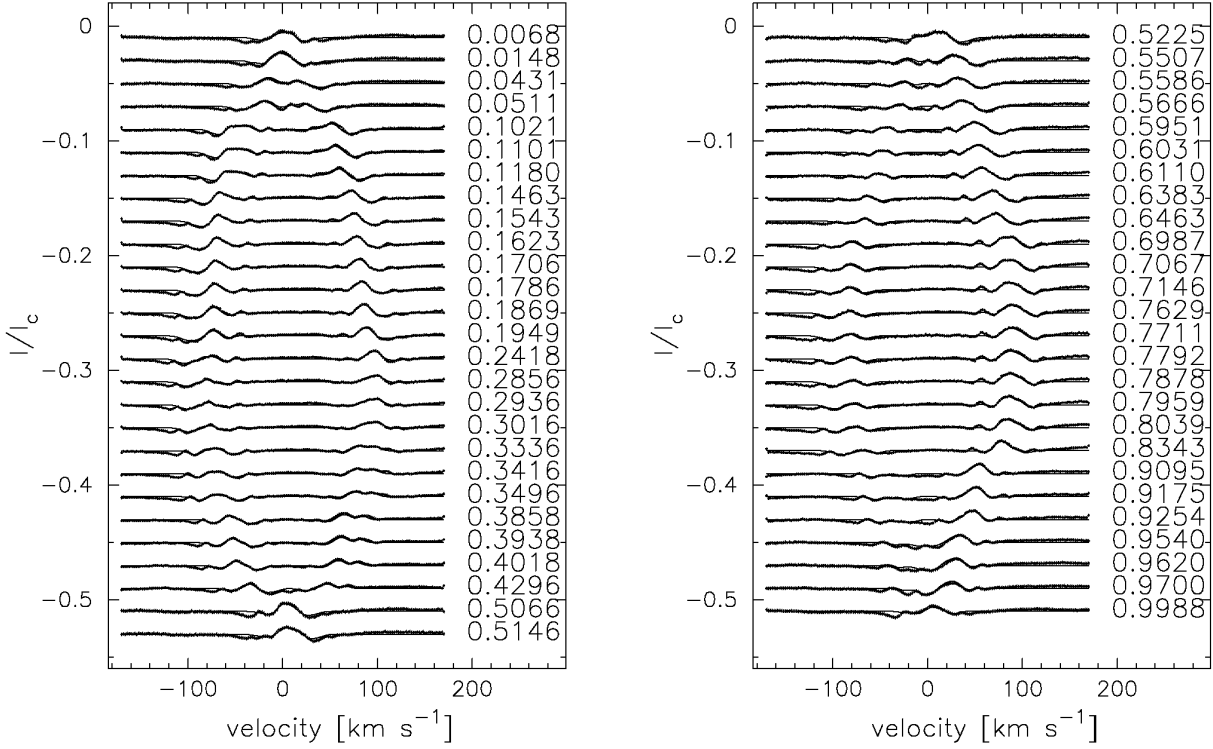}
 \end{center}
\caption[Stokes I fits]{ONLINE ONLY: The 2007 Stokes I LSD profiles and fits are shown after subtraction of the model immaculate (unspotted) stellar profiles. The solid line is the maximum entropy regularised fit we obtain to the data from Doppler imaging. Profiles are ordered by orbital phase (listed to the right of each profile). For reference, the primary star is the profile seen generally at negative velocities for phases $\phi=0.0-0.5$. The residuals in the centre of the stellar profiles are starspots and are mostly well modelled. We note however that poor agreement is found at the extremities of the rotational profiles of both stars. This is similar to that observed for the contact binary AE Phe by \cite{barnes04}. Here the authors attributed the mis-match to inadequate modelling of centre-to-limb brightness variation. A more sophisticated treatment of limb darkening beyond the linear limb darkening used here is required.}
\protect\label{fig:fitspot}
\end{figure*}

\end{document}